\documentclass[fleqn,usenatbib]{mnras}

\usepackage{newtxtext,newtxmath}

\usepackage[T1]{fontenc}

\DeclareRobustCommand{\VAN}[3]{#2}
\let\VANthebibliography\thebibliography
\def\thebibliography{\DeclareRobustCommand{\VAN}[3]{##3}\VANthebibliography}

\usepackage{graphicx}	
\usepackage{amsmath}	

\def\Lw{\textit{Lightweaver}}

\title[Mixing-induced cooling signatures]{Observational signatures of mixing-induced cooling in the Kelvin-Helmholtz instability}

\author[B. Snow et al.]{
B. Snow,$^{1}$\thanks{E-mail: b.snow@exeter.ac.uk}
C. Osborne, $^2$
A. S. Hillier,$^{1}$
\\
$^{1}$University of Exeter, Exeter, EX4 4QF, UK \\
$^{2}$SUPA School of Physics \& Astronomy, University of Glasgow, UK
}

\date{Accepted XXX. Received YYY; in original form ZZZ}

\pubyear{2015}

\begin{document}
\label{firstpage}
\pagerange{\pageref{firstpage}--\pageref{lastpage}}
\maketitle

\begin{abstract}
Cool ($\approx 10^4$K), dense material permeates the hot ($\approx 10^6$K), tenuous solar corona in form of coronal condensations, for example prominences and coronal rain. 
As the solar atmosphere evolves, turbulence can drive mixing between the condensations and the surrounding corona, with the mixing layer exhibiting an enhancement in emission from intermediate temperature ($\approx10^5$K) spectral lines, which is often attributed to turbulent heating within the mixing layer. 
However, radiative cooling is highly efficient at intermediate temperatures and numerical simulations have shown that radiative cooling can far exceed turbulent heating in prominence-corona mixing scenarios. As such the mixing layer can have a net loss of thermal energy, i.e., the mixing layer is cooling rather than heating.   
Here, we investigate the observational signatures of cooling processes in Kelvin-Helmholtz mixing between a prominence thread and the surrounding solar corona through 2D numerical simulations. 
Optically thin emission is synthesised for Si\,\textsc{iv}, along with optically thick emission for H$\alpha$, Ca\,\textsc{ii}\,K and Mg\,\textsc{ii}\,h using \Lw. The Mg\,\textsc{ii}\,h probes the turbulent mixing layer, whereas H$\alpha$ and Ca\,\textsc{ii}\,K form within the thread and along its boundary respectively. 
As the mixing evolves, intermediate temperatures form leading to an increase in Si\,\textsc{iv} emission, which coincides with increased radiative losses. The simulation is dominated by cooling in the mixing layer, rather than turbulent heating, and yet enhanced emission in warm lines is produced.
As such, an observational signature of decreased emission in cooler lines and increased emission in hotter lines may be a signature of mixing, rather than an implication of heating.
\end{abstract}




\section{Introduction}

Cool, dense material is often observed {within} in the hot, tenuous solar corona in the form of solar prominences/{filaments} \citep{Parenti2014}, coronal rain \citep{Antolin2020}, and spicules \citep{Beckers1968,Pereira2019,Hinode2019}. The density within these features can be several orders of magnitude larger than the surrounding material, in addition to the temperature being orders of magnitude lower. The Prominence-Corona Transition Region (PCTR) that forms between these layers as a result of turbulent mixing contains material at intermediate temperatures of around $10^5$K. {The PCTR is essential in the formation of spectra \citep{Labrosse2004,Heinzel2015}.}


There are a number of studies that show decreased emission in cool ($10^4$K) spectral lines, alongside a co-temporal increase in emission at warm lines ($10^5$K) in both prominence threads \citep{Okamoto2015} and spicules \citep{dePontieu2007,Beckers1972,Pereira2014}. The explanation given by \cite{Okamoto2015} is that the transverse oscillations of the prominence thread excite the Kelvin-Helmholtz Instability (KHI). Resonant absorption dissipates the kinetic energy within the mixing layer, with some of the dissipated energy contributing towards heating \citep{Antolin2015}. However, the mixing layer that forms between the prominence thread and corona has a temperature around $10^5$K. Solar material is subject to very efficient radiative losses and strong cooling that far exceeds the heating due to turbulent dissipation \citep{Hillier2019,Hillier2023}. 
As such, alternative explanations must be explored for the observational signatures of condensations fading in cool lines and appearing in hotter lines.


The fundamental mixing process provides an alternative interpretation of the co-temporal fading of cool emission and emergence of warm material.  
In the mixing layer, dense cool material is mixed with tenuous hot material, leading to intermediate temperatures and densities. The cool material is replenished in a more diffuse phase, leading to a net decrease in emission in cool spectral lines \citep{Hillier2023}. Also, since intermediate temperatures are forming, hotter transitions-region spectral lines will have enhanced emission. Therefore, the observed signatures may be obtained from the fundamental mixing process, without requiring heating of the material. In fact, due to the strong cooling that can occur in the mixing layer such a signature may be possible to obtain in a case where there is a net decrease in thermal energy.



In this paper, the observational signatures of condensation-corona mixing are studied in a cooling-dominated numerical simulation of the Kelvin-Helmholtz Instability (KHI). The simulation features strong radiative losses and has a net-decrease in thermal energy with time, i.e., the cooling is far more effective than the turbulent heating. Intensity is synthesised for both optically-thin and optically-thick spectral lines using a combination of CHIANTI v9 \citep{Dere1997,Dere2019} and \Lw{} \citep{Osborne2021}. The results indicate that the mixing process itself may be responsible for the observed behaviour of prominence fading, without requiring heating to occur.


\section{Methods}


The hypothesis of this paper is that heating-like observational signatures can be obtained as a result of the mixing between prominence and coronal material due to the enhanced radiative losses that occur within the mixing layer. To test this hypothesis, a numerical simulation is performed of the mixing between a falling prominence thread and the surrounding solar corona. 
The falling thread is assumed to be 100 times more dense than the surrounding corona and, since it is also assumed that initially there is constant pressure, the thread is also 100 times cooler than the corona. Typical values are chosen of a solar corona that is $10^6$K, thus the thread is $10^4$K. A reference electron number density in the solar corona is chosen as $n_{\rm norm}=10^{15}\mbox{~m}^{-3}$.


\begin{table}
    \centering
    \caption{Normalisation constants and the corresponding dimensional properties of the corona and thread}
    \begin{tabular}{c|c|c|c}
                  & Normalisation value & Corona & Thread \\ \hline 
        {$\hat{v}$ [km/s]} & {$166$} & 0 & {16.6}\\
        $\hat{T}$ [K] & $10^6$ & $10^6$ & $10^4$\\
        $\hat{n_e}$ [m$^{-3}$] & $10^{15}$ & $10^{15}$ & $10^{17}$\\
        $\hat{x}$ [Mm] & 1 & & \\
        $\hat{t}[s]$ & $\hat{x}/\hat{v}\approx 6$ & &
    \end{tabular}
    \label{tab:constants}
\end{table}

\subsection{Numerical Simulation}

\begin{figure} 
	\includegraphics[width=\columnwidth]{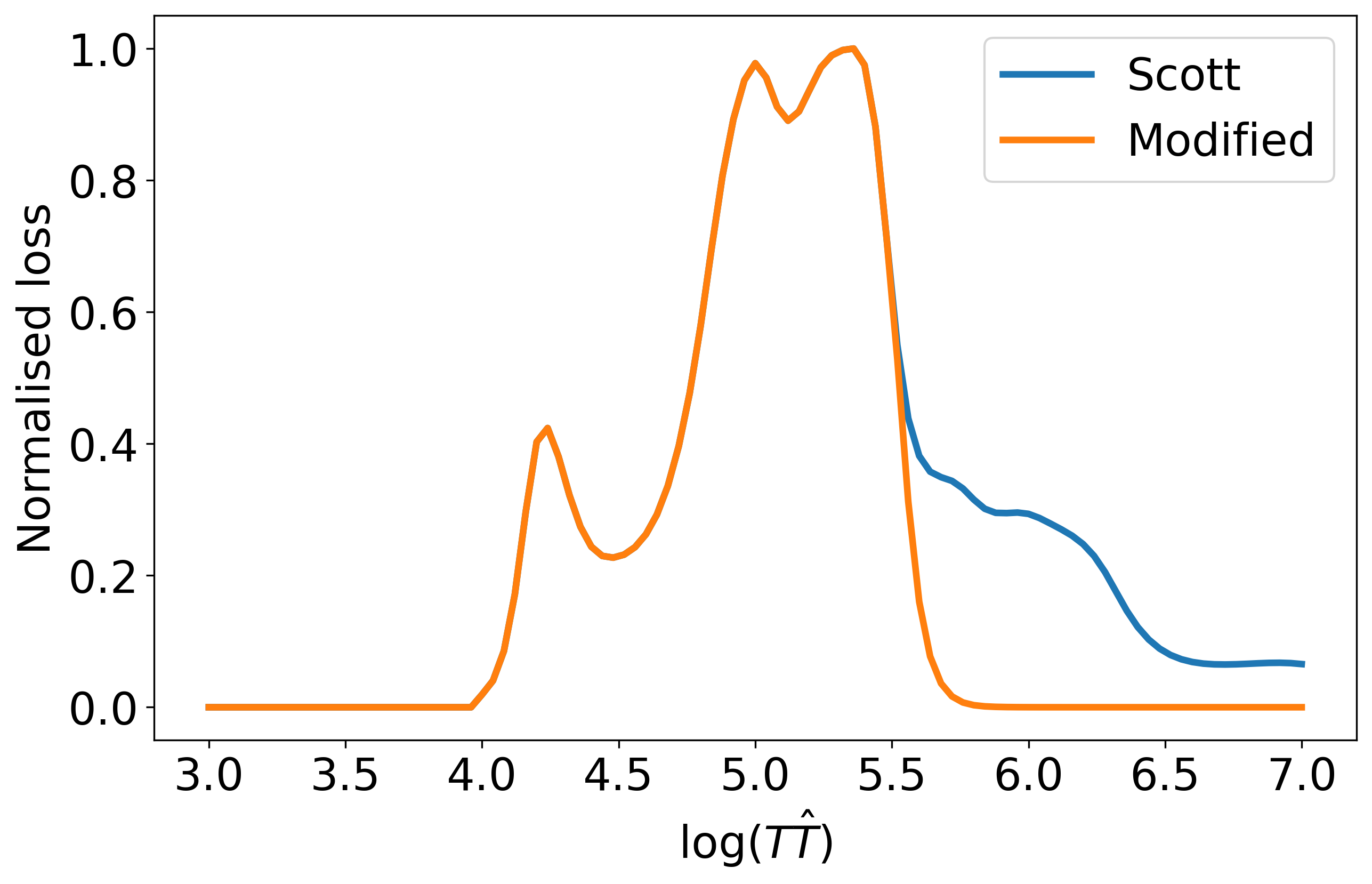}
    \caption{Radiative loss profile from CHIANTI using the abundances of \citep{Scott2015} (blue), and the modified loss function used in the simulation (orange). The quantity $T\hat{T}$ is the dimensional temperature in Kelvin.}
    \label{fig:lossprofile}
\end{figure}


Numerical simulations are performed using the (P\underline{I}P) code \citep{Hillier2016}, to evolve the optically-thin radiative MHD equations in non-dimensional form. 
Specifically, the following equations are evolved:
\begin{gather}
\frac{\partial \rho}{\partial t} + \nabla \cdot (\rho \textbf{v}) = 0, \\
\frac{\partial}{\partial t} (\rho \textbf{v})+ \nabla \cdot \left( \rho \textbf{v} \textbf{v} + P \textbf{I} - \textbf{B B} + \frac{\textbf{B}^2}{2} \textbf{I} \right) = 0,\\
\frac{\partial}{\partial t} \left( e + \frac{\textbf{B}^2}{2} \right) + \nabla \cdot \left[ \textbf{v} ( e + P) -  (\textbf{v} \times \textbf{B}) \times \textbf{B} \right]  =  - \rho ^2 \Lambda(T\hat{T}), \\
\frac{\partial \textbf{B}}{\partial t} - \nabla \times (\textbf{v} \times \textbf{B}) = 0, \\
\nabla \cdot \textbf{B} =0, \\
e = \frac{P}{\gamma -1} + \frac{1}{2} \rho v ^2, 
\end{gather}
for density $\rho$, velocity $\textbf{v}=\left[ v_x,v_y,v_z \right]$,  magnetic field $\textbf{B}=\left[ B_x,B_y,B_z \right]$ and pressure $P$. The temperature is assumed to obey the non-dimensional ideal gas law $T=\frac{\gamma P}{\rho}$. The loss function $\Lambda$ depends on the dimensional temperature $T\hat{T}$ which is the simulation temperature multiplied by the normalisation temperature $\hat{T}$. There is no explicit viscosity or resistivity, however some numerical dissipation occurs due to the underlying numerical methods.

The system is non-dimensionalised using the reference values of $\hat{T}=10^6$K in the solar corona, and a reference electron number density inside the thread of $n_e=10^{15}\mbox{~m}^{-3}$, with both temperature and density varying in the domain. 
A lengthscale is chosen as $L_{\rm norm}=1$Mm. The coronal sound speed is normalised based on the temperature as $c_{s,\rm norm}=\sqrt{\frac{\gamma R_g \hat{T}}{M}} \approx 166 \sqrt{\hat{T}}{m/s = 166}${~km/s}, where the molar gas constant $R_g\approx 8.314\mbox{~m}^2\mbox{~kg}\mbox{~s}^{-2}\mbox{~K}^{-1}\mbox{~mol}^{-1}$ and $M=0.5 \times 10^{-3} \mbox{~kg}\mbox{~mol}^{-1}$ is chosen based on the mean mass of atomic species present in the solar corona \citep{Priest1982}. The timescale is then $t(s)=\hat{v}/\hat{x}\approx 6$, which is the cadence of the simulation output. The normalisation constants for the system are given in Table \ref{tab:constants}. 

\subsection{Radiative losses}
Energy is removed from the system due to optically thin radiative losses, which are assumed to be of the form $\rho^2 \Lambda (T\hat{T})$, where $\Lambda(T\hat{T})$ is a temperature dependent function shown in Figure \ref{fig:lossprofile} using the dimensional temperature $T\hat{T}$. The loss function is created using CHIANTI v9 \citep{Dere1997,Dere2019} with the default abundance file. It is beneficial numerically to have zero losses either side of the interface initially, such that the losses only apply to the mixing layer. As such, the loss profile is modified according to 
\begin{gather}
    \Lambda (T\hat{T})= 
\begin{cases}
    0,& \text{if } \log_{10}T\hat{T} < 4\\
    \Lambda_0 (T\hat{T}),& \text{if }  4 <\log_{10}T\hat{T} < 5.5\\
    \Lambda_0 (T\hat{T})(1.0-\tanh\left( \left( \frac{\log_{10}(T\hat{T})-5.5}{0.4}\right)^2 \right),  & \text{otherwise}
\end{cases}
\end{gather}
such that below $T\hat{T}=10^{5.5}$K, the temperature profile is identical to the default loss profile. Above this temperature, the losses are gradually reduced to zero. The implicit assumption here is that some anomalous heating is applied above $T\hat{T}=10^{5.5}$K that is sufficient to balance the coronal losses. It is also assumed that there is no cooling below $T\hat{T}<10^{4}$K. The cooling curve is shown in Figure \ref{fig:lossprofile}. 

\subsection{Initial Conditions}
The initial numerical set up is similar to the model presented in \citet{Hillier2023} however, here, a realistic radiative loss curve is used, with the timescale for radiative cooling controlled by the free parameter $\tau_{rad}=10^{-2}$, i.e., for a density of 1, losses occur on timescales of 100 time units (i.e., 600 s) at the peak of the cooling curve. The losses are calculated at each timestep according to the local density and temperature. As such, losses can occur on far shorter timescales when high densities form at intermediate temperatures.

The pressure is constant across the initial domain as $P=1/\gamma$, with the density set according to
\begin{equation}
\rho=\begin{cases}\begin{array}{lr}
100 & \text{for } y>-1,\\
1 & \text{for } y<-1.\end{array}
\end{cases}
\end{equation}
corresponding to a dimensional temperature of $T_{\rm corona}=10^6$K and $T_{\rm thread}=10^4$K, using the normalisation parameters in Table \ref{tab:constants}. The magnetic field is uniform in the out-of-plane ($z-$)direction with the initial coronal plasma-$\beta=0.05$.

The initial flow is set to be in the $x$ direction of
\begin{equation}
v_x=\begin{cases}\begin{array}{lr}
0.1 & \text{for } y>-1,\\
0 & \text{for } y<-1,
\end{array}
\end{cases}
\end{equation}
to mimic a dense thread falling through the tenuous solar corona. A small random $v_y$ perturbation with a maximum magnitude of 0.01 centred on the interface is included to trigger the instability. 

\subsection{Simulation Evolution}
\begin{figure}
	\includegraphics[width=\linewidth]{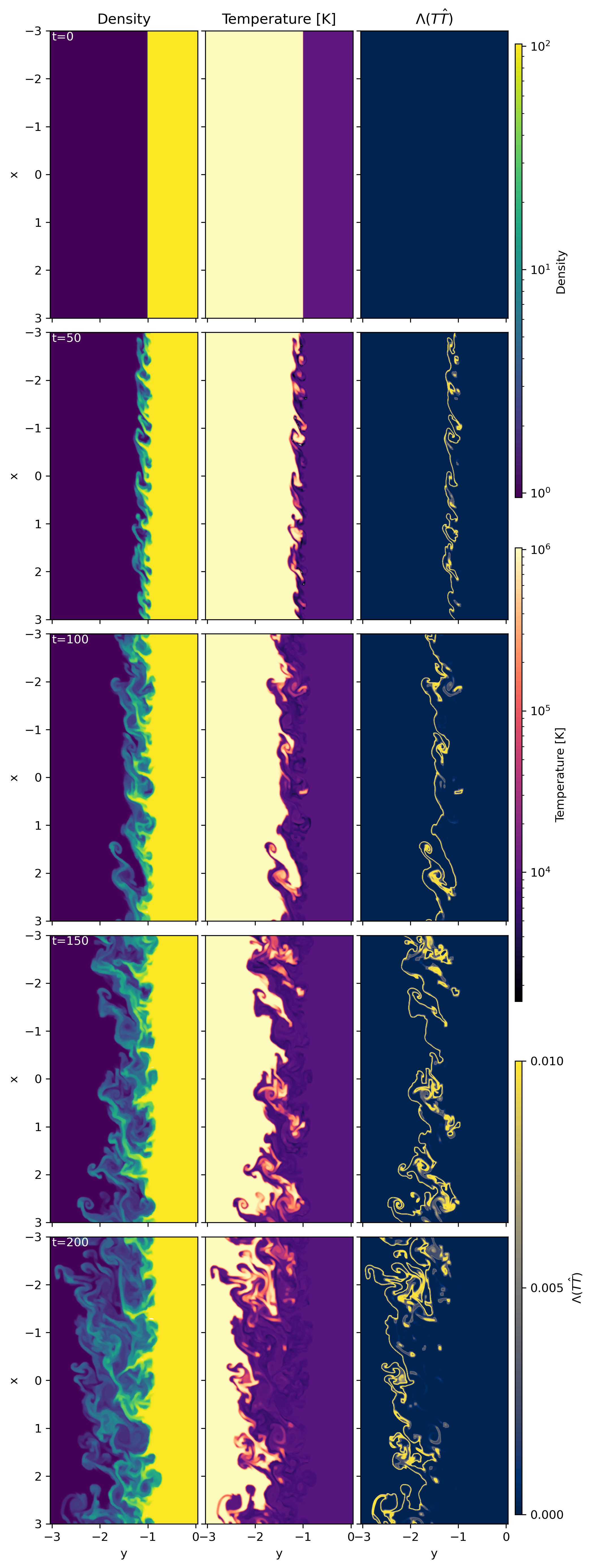}
    \caption{Evolution of the KHI with radiative losses showing the density (top row), temperature (middle row) and radiative losses $\Lambda(T)$ (lower row). Time, density and loss rate are in simulation units. Temperature is given in Kelvin.}
    \label{fig:simevo}
\end{figure}

The evolution of the density and temperature in the simulation is shown in Figure \ref{fig:simevo}. In the mixing layer, the mixed temperature becomes subject to strong radiative losses and thus efficiently cools. As such, the temperature interface between the two layers becomes reasonably sharp, as discussed in detail in \citet{Hillier2023}. The thermal energy of the system reduces through time due to the radiative losses, and the energy lost due to this cooling is far greater than the turbulent heating provided by numerical dissipation. As such, the mixing layer acts to \textbf{cool} the system. {Note that this is true for any cooling curve where the maximum cooling rate occurs within the mixing layer \citep[e.g.,][where an idealised cooling curve is used]{Hillier2023}.}

\section{Synthetic observables}

\subsection{Observing configuration}

\begin{figure}
    \includegraphics[width=\columnwidth]{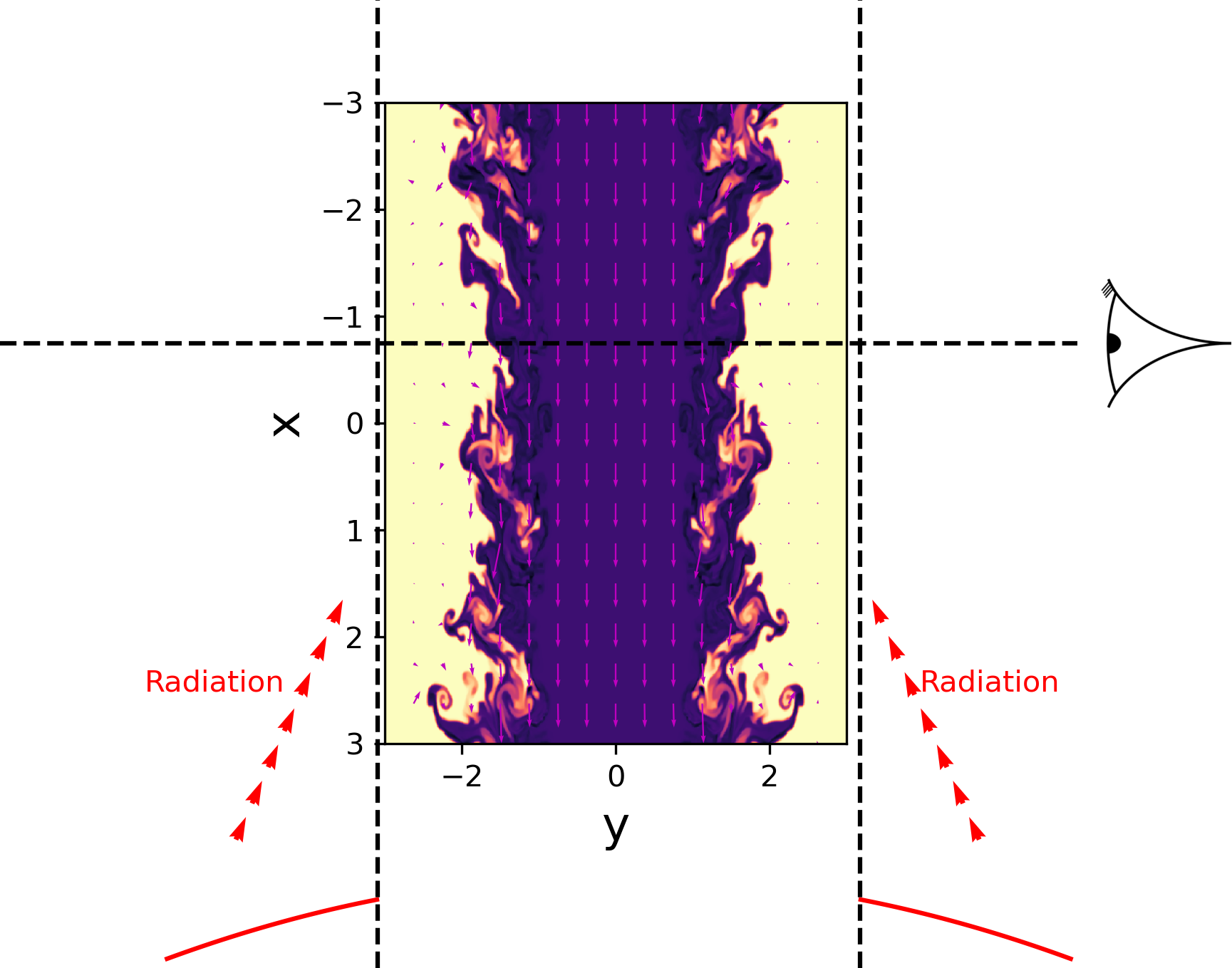}
    \caption{Observing configuration for the synthetic data. The simulation data is mirrored around $y=0$. Line-of-sight is indicated by the dashed line. Background colour is the temperature in kelvin. Velocity vector is overlaid as arrows. The simulation snapshot shown here is at dimensionless time $t=150$, corresponding to dimensional time $\hat{t}\approx 900$s. The curved red line indicates the source of radiation.}
    \label{fig:obs_setup}
\end{figure}

For the optically thick line synthesis, well defined boundaries have to be imposed to treat the radiation. Here, this is performed by mirroring the numerical data to mimic observing a downflowing thread that has coronal material either side, as shown in Figure \ref{fig:obs_setup}. The line-of-sight is perpendicular to the thread, as indicated by the black dashed line in Figure \ref{fig:obs_setup}. A combination of optically-thin and optically-thick emission synthesis is applied depending on the line.

\subsection{Optically thin emission}


A simple way to estimate observable emission is to assume the medium is optically thin and thus the intensity is calculated as
\begin{gather}
    I=\int \rho ^2 G(T\hat{T}) dl
\end{gather}
which is integrated over some line-of-sight $l$. The contribution function $G(T\hat{T})$ depends on the local (dimensional) temperature and is synthesised using CHIANTI v9 \citep{Dere1997,Dere2019}. The optically-thin approach is quick to synthesise since the emission depends on the local properties only, and light emitted at any point along the line of sight is considered to freely propagate towards the observer. In this paper, the Si\,\textsc{iv} line is used as an example of an optically-thin line that forms around transition region temperatures, namely, between approximately $4.4\times 10^4 \leq T\hat{T} \leq 3.3\times 10^5$\,K. 

\subsection{Optically thick emission approach}\label{Sec:OpticallyThickSynth}



For optically-thick emission, a different approach is required to solve the radiative transfer problem, which is performed here using the open-source \Lw{} framework \citep{Osborne2021}. 
\Lw{} computes atomic excitation and ionisation in plasma where these depart from local thermodynamic equilibrium {(i.e., the so-called non-LTE state is calculated)}, considering collisional and radiative effects in atomic lines and continua in detail.
Here the model is set up as per the prominence projection and boundary conditions described in \citet{Jenkins2023}.
Each line-of-sight row (as shown in Figure \ref{fig:obs_setup}) is treated independently in a plane-parallel manner.
On each end of this column we employ boundary conditions that account for the incident irradiation from the solar disk, taking into account the effects of limb darkening.
The solar irradiation is computed from the semi-empirical quiet sun FAL C model \citep{Fontenla1993}, and is tabulated for different viewing angles under a plane parallel approximation.

\Lw{} then computes the (steady-state) statistical equilibrium solution for atomic excitation and ionisation along with the associated line profiles.
This is achieved via the multi-level accelerated lambda iteration approach of \citet{Rybicki1992}, along with the hybrid partial frequency redistribution (PRD) treatment of \citet{Leenaarts2012} implemented with the method of \citet{Uitenbroek2001}.
PRD is considered for the resonance lines investigated, namely Ca\,\textsc{ii} K and Mg\,\textsc{ii} h.

The optically thick synthesis requires significant CPU time and convergence is not guaranteed. Locations that did not converge are treated as missing data in this paper and are not included in any analysis. 

\subsection{Formation locations}

\begin{figure}
    \centering
    \includegraphics[width=0.99\linewidth]{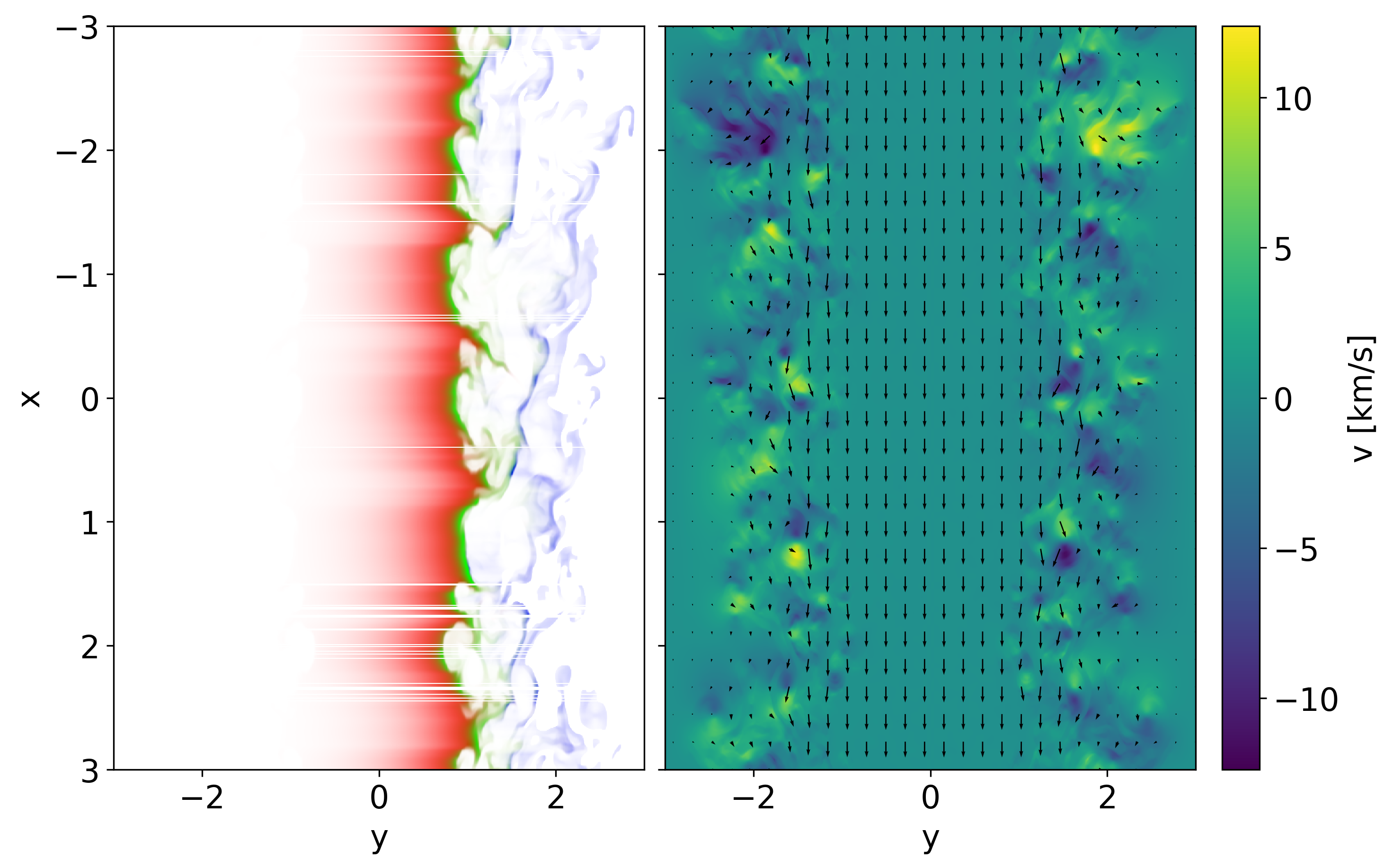}
    \caption{(left) Optically-thick emission locations for the line-core of H$\alpha$ (red), Ca\,\textsc{ii}\,K  (green) and Mg\,\textsc{ii}\,h (blue) at time $\hat{t}=1200$s (corresponding to simulation time $t=200$ in Figure \ref{fig:simevo}). Unconverged locations are denoted by horizontal white stripes and are not used in any analysis (right) $v_y$ velocity colourmap from the simulation with arrows denoting velocity vectors. $v_y$ is the velocity component aligned with the line-of-sight for the synthetic observables.}
    \label{fig:combined_formation}
\end{figure}

\begin{figure}
    \centering
    \includegraphics[width=0.99\linewidth]{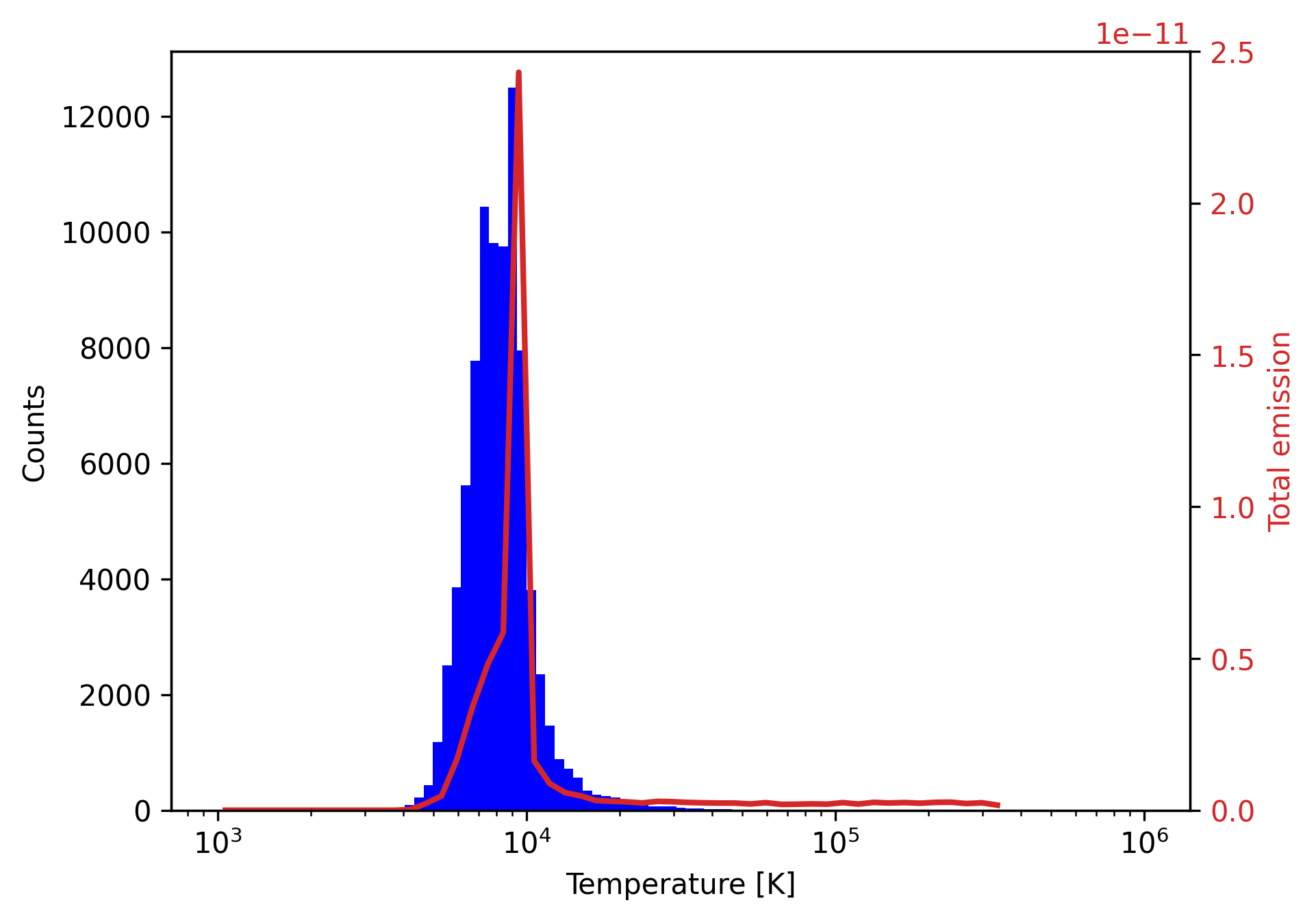}
    \caption{Histogram of the temperature at locations where the Mg\,\textsc{ii}\,h line forms. Overplotted is the total emission from each temperature band.}
    \label{fig:mg_temperature}
\end{figure}

The formation of the spectral lines studied here sample different locations in the simulation, as shown in Figure \ref{fig:combined_formation} for dimensional time $\hat{t}(s)\approx 1200 s$. 
Defining the spatially-, angularly-, and wavelength-varying emissivity of the medium as $\eta$, and the spatially-, angularly-, and wavelength-varying opacity as $\chi$, the locations that are contributing towards the emission is each spectral line can be identified. 
The intensity contribution function shown here is the integrand of the radiative transfer equation to the line core, highlighting the regions primarily responsible for the emergent spectrum. 

As shown in Figure \ref{fig:combined_formation}, H$\alpha$ predominantly forms in the central (effectively unmixed) region, Ca\,\textsc{ii}\,K forms at the edge the mixing layer (and thus is also effectively unmixed) and thus the emission is primarily sampling the thread. The Mg\,\textsc{ii}\,h forms within the mixing layer itself, and therefore samples entrained material. Note that the line-of-sight goes from the right of the image towards the left and thus the contribution towards emission comes broadly from the right half of the domain. Given the different formation locations of the spectral lines, one would expect different properties to be present in the emission. The line-of-sight velocity component $v_y$ is shown across the domain in Figure \ref{fig:combined_formation}(right) and shows where the mixing is occurring. As such, one would expect that the Doppler velocity in the Mg\,\textsc{ii}\,h line is significantly larger than the velocities observed in the Ca\,\textsc{ii}\,K and H-$\alpha$ lines, which can be analysed by studying the asymmetry of the spectral emission.    


The Mg\,\textsc{ii}\,h line formation location has a range of different temperatures, shown in the histogram in Figure \ref{fig:mg_temperature}, with the peak emission coming from material around $T=10^4$K. This is roughly the temperature of the initial thread, however, from the formation locations in Figure \ref{fig:combined_formation}, one can see that the Mg\,\textsc{ii}\,h line-core is forming mostly within the mixing layer. The strong radiative losses in the mixing layer cause the mixing material to rapidly cool and thus the Mg\,\textsc{ii}\,h formation locations are relatively thin bands. 

\begin{figure*}
    \centering
    \includegraphics[width=0.32\linewidth]{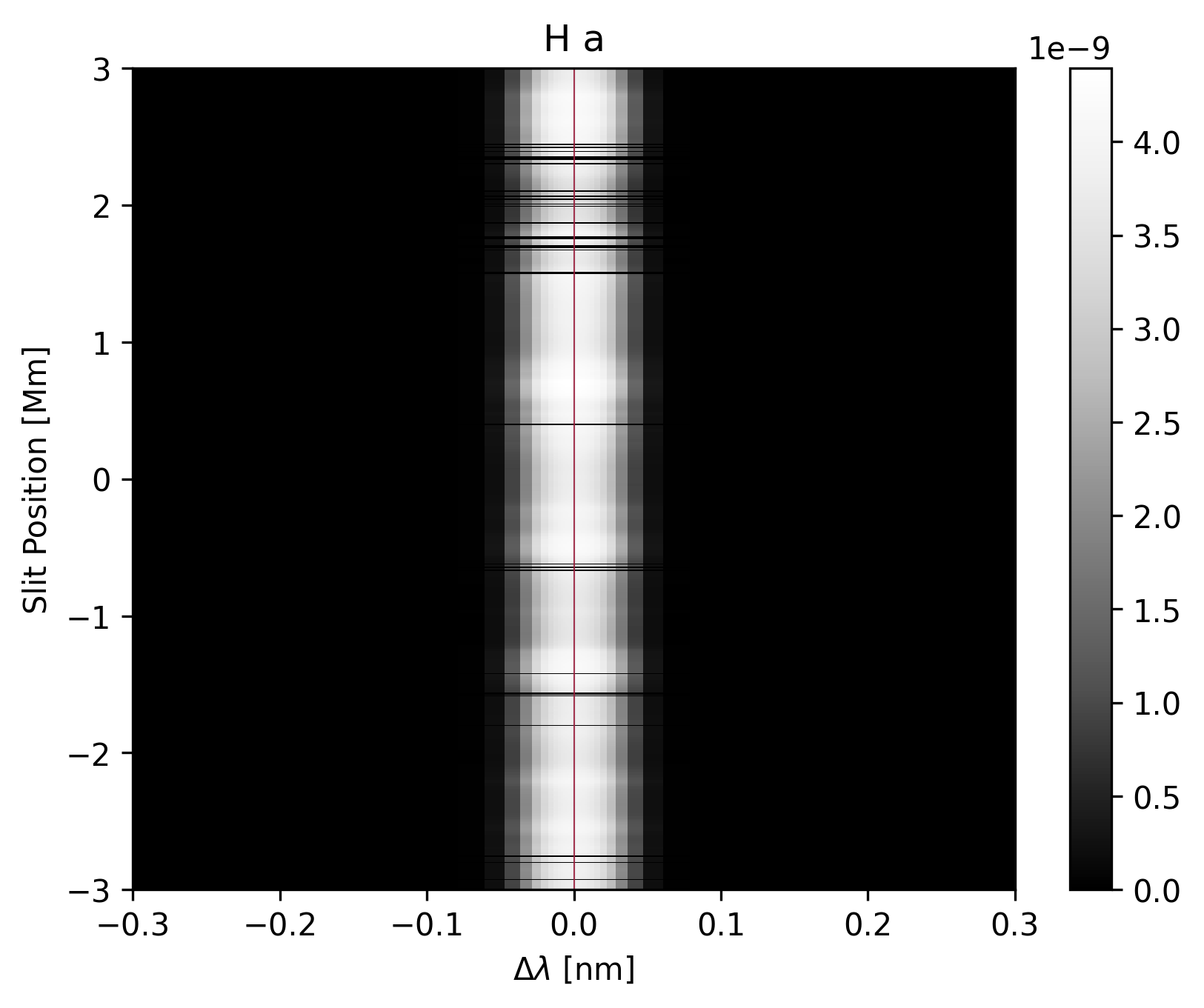}
    \includegraphics[width=0.32\linewidth]{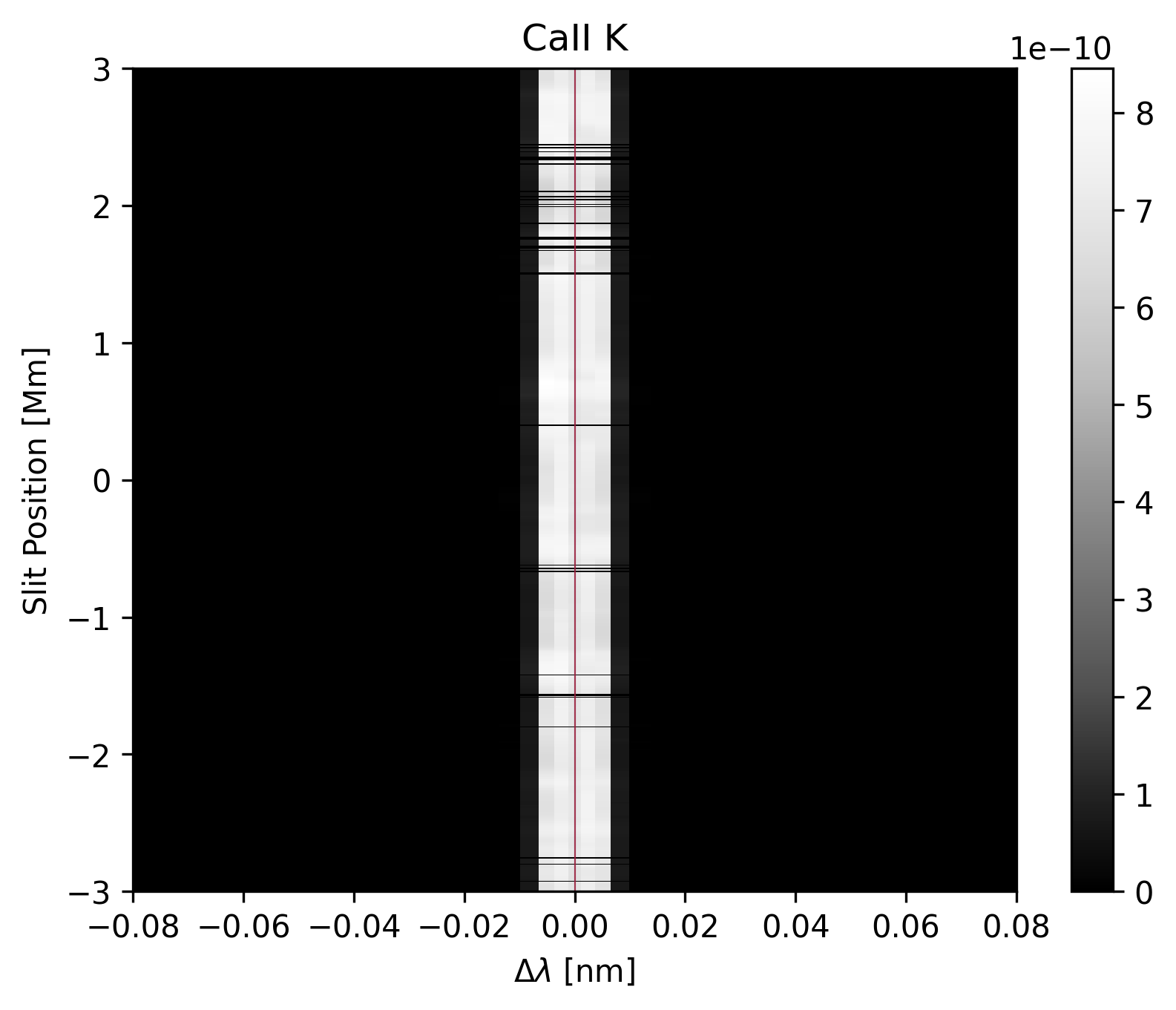}
    \includegraphics[width=0.32\linewidth]{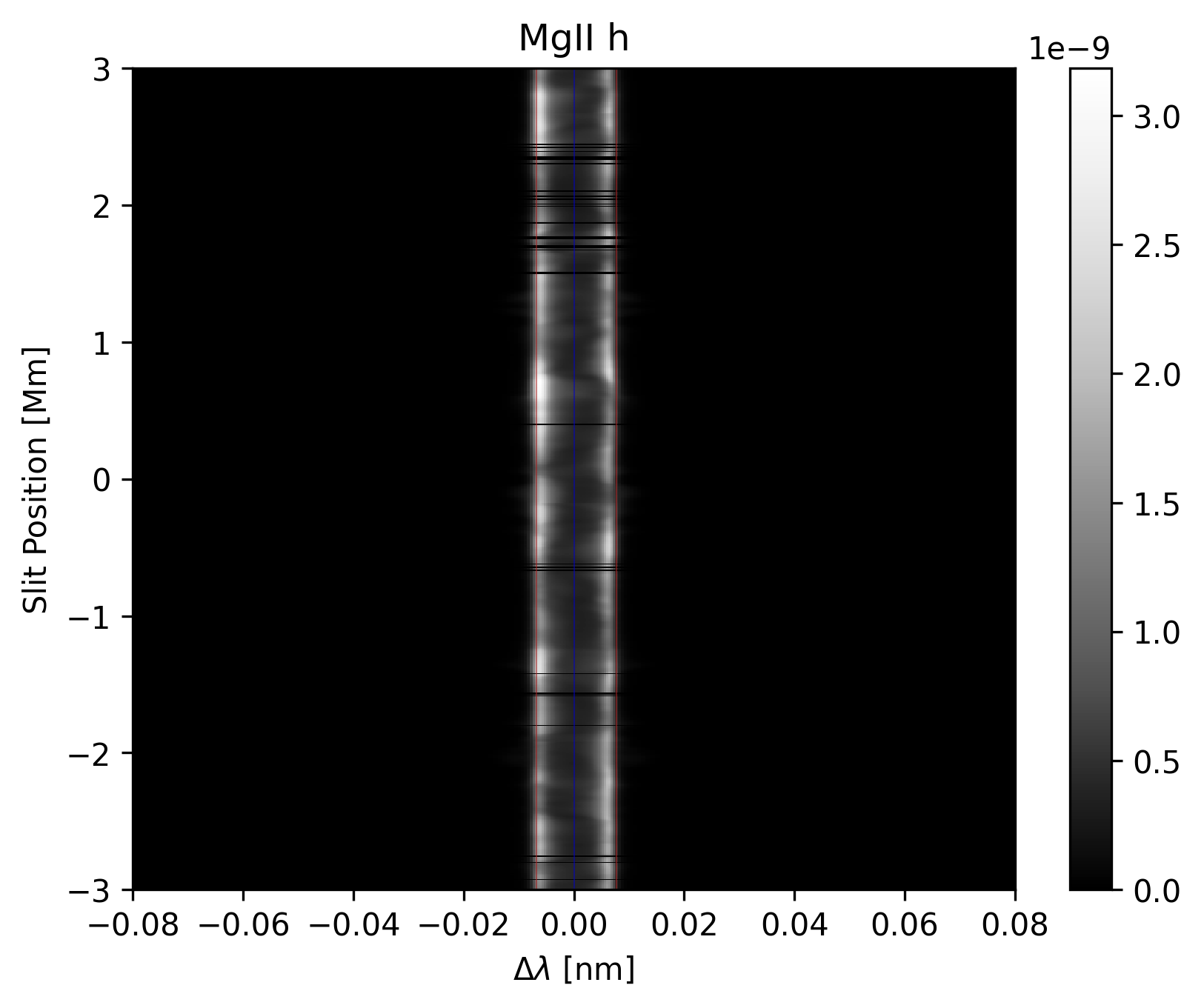}
    \caption{Synthesised spectral emission from Halpha, Ca\,\textsc{ii}\,K, Mg\,\textsc{ii}\,h throughout the domain at time $\hat{t}=1200$s (corresponding to simulation time $t=200$). Unconverged data is shown as the black horizontal lines.}
    \label{fig:slitspectra}
\end{figure*}

\begin{figure*}
    \centering
    \includegraphics[width=0.32\linewidth]{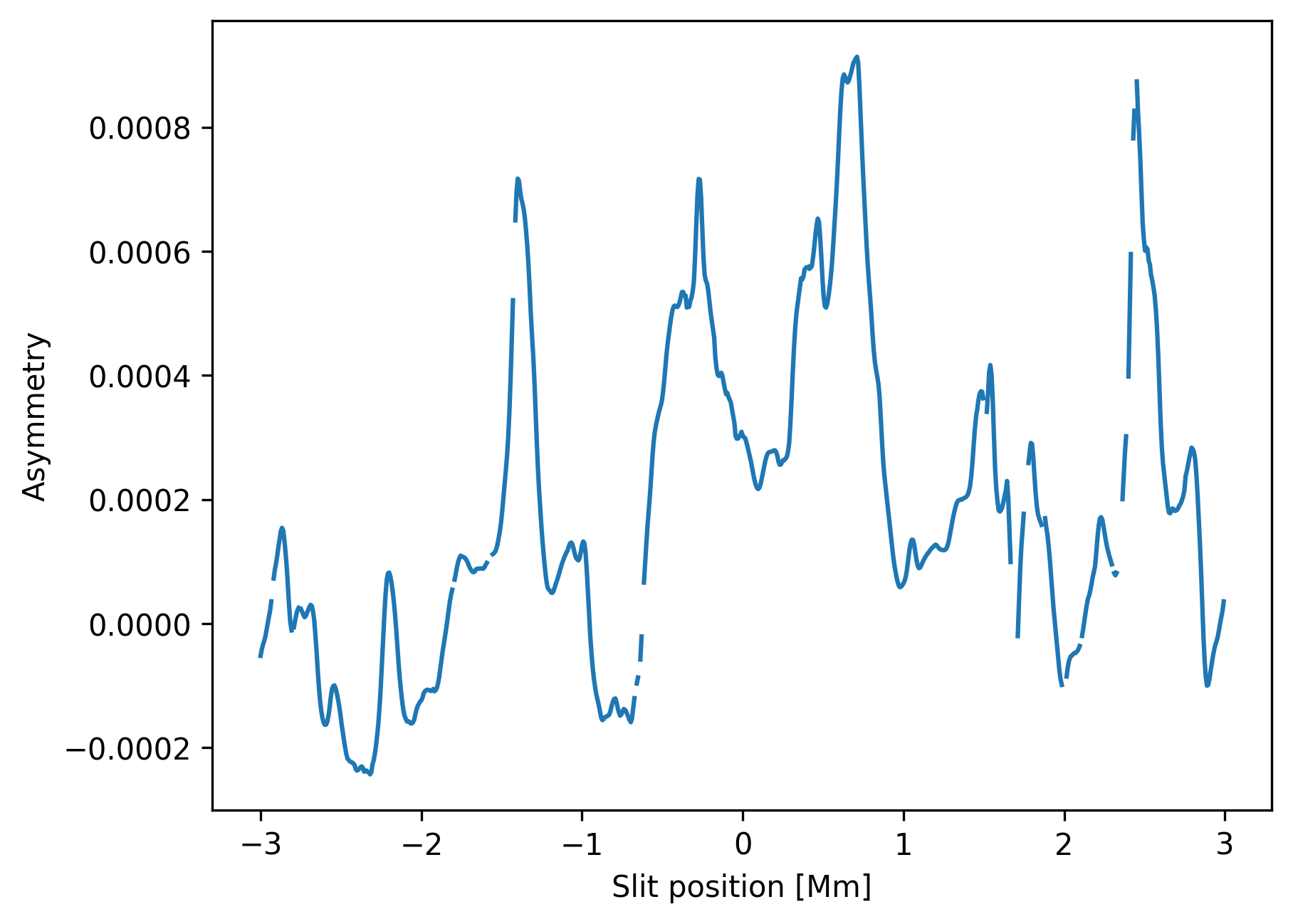}
    \includegraphics[width=0.32\linewidth]{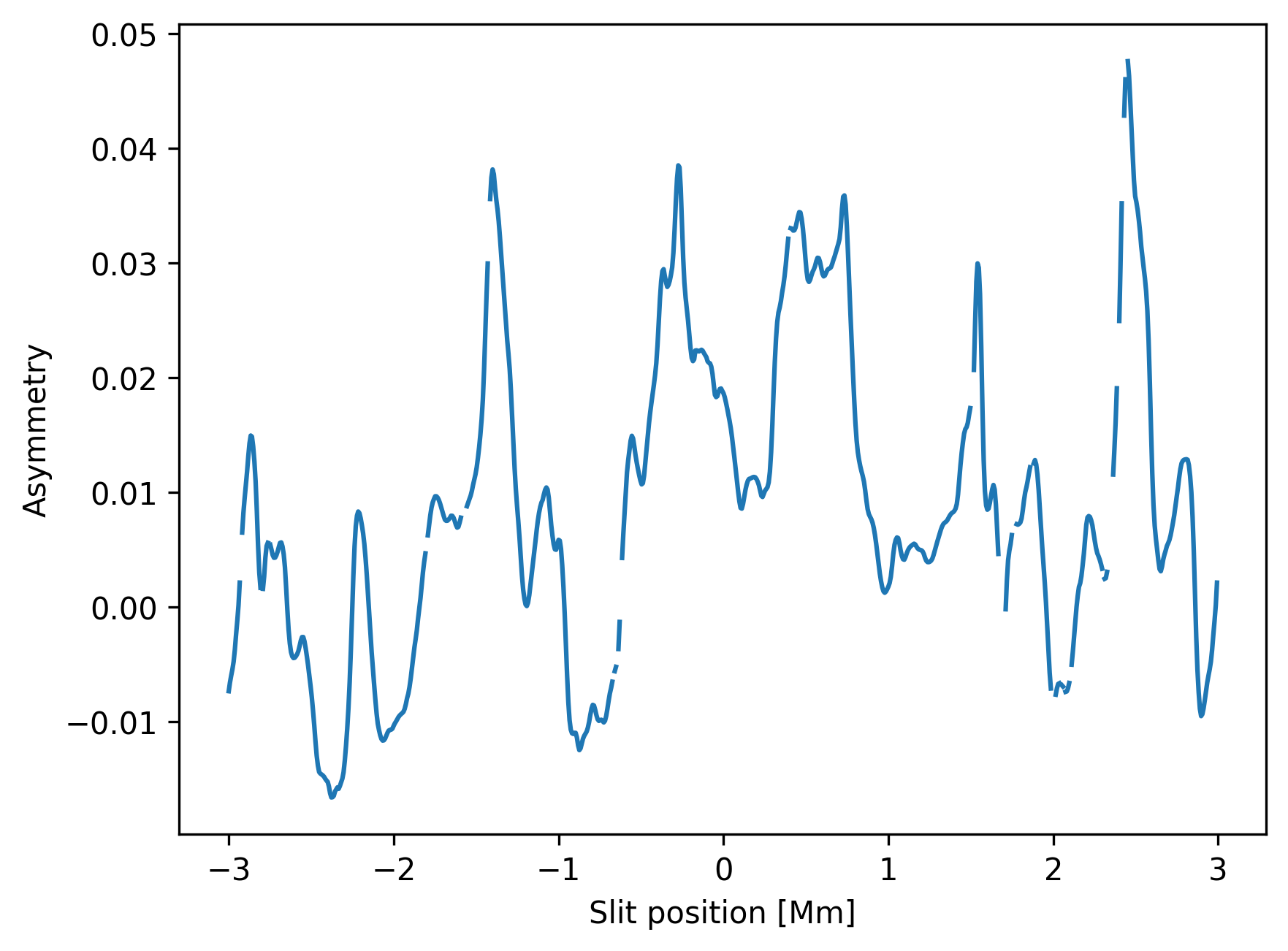}
    \includegraphics[width=0.32\linewidth]{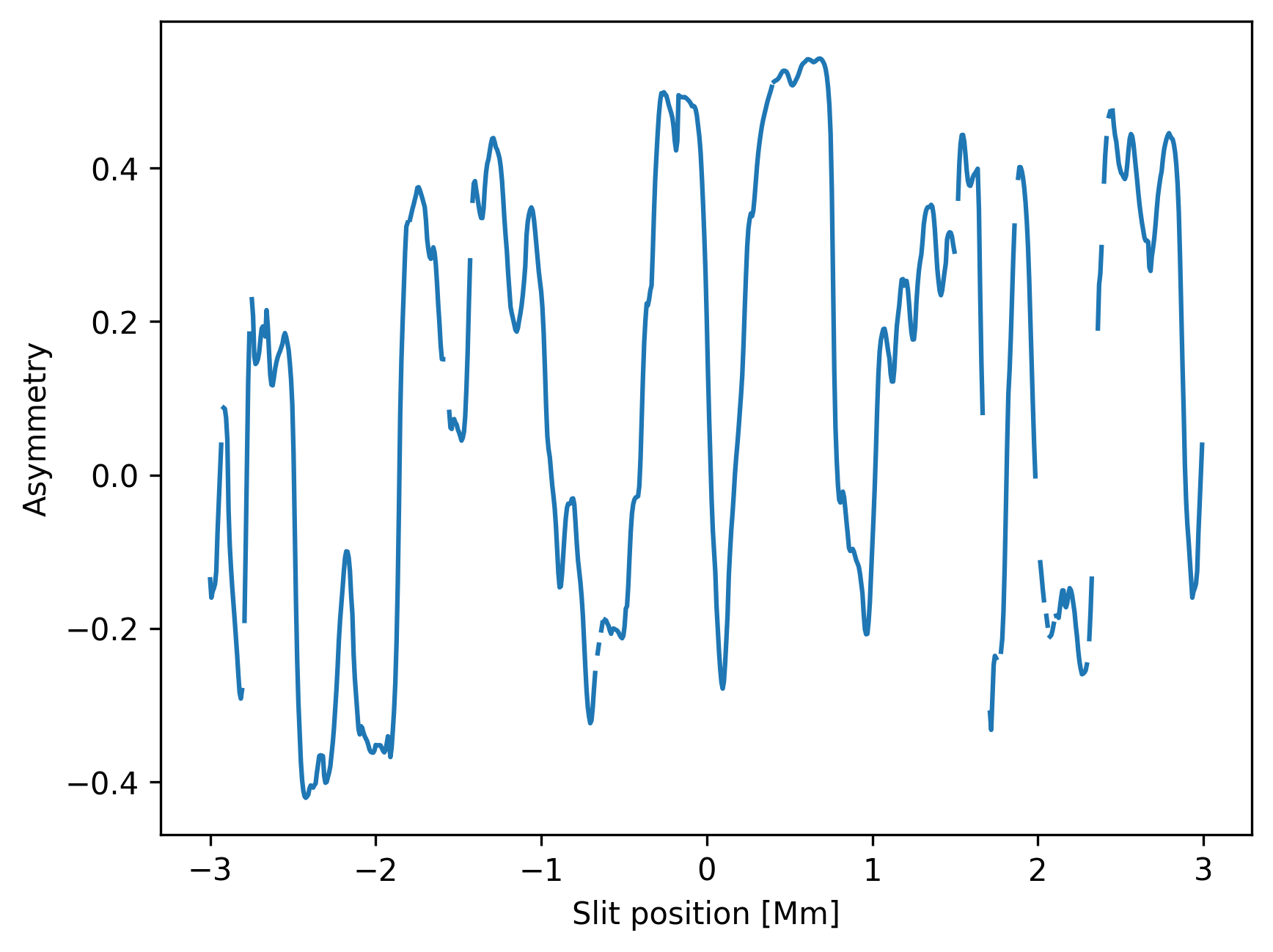}
    \caption{Asymmetry of the synthesised spectral emission from H$\alpha$ (left), Ca\,\textsc{ii}\,K (centre), Mg\,\textsc{ii}\,h (right) at time $\hat{t}=1200$s (corresponding to simulation time $t=200$). Unconverged data is treated as missing data.}
    \label{fig:spectraAsym}
\end{figure*}

\begin{figure*}
    \centering
    \includegraphics[width=0.32\linewidth]{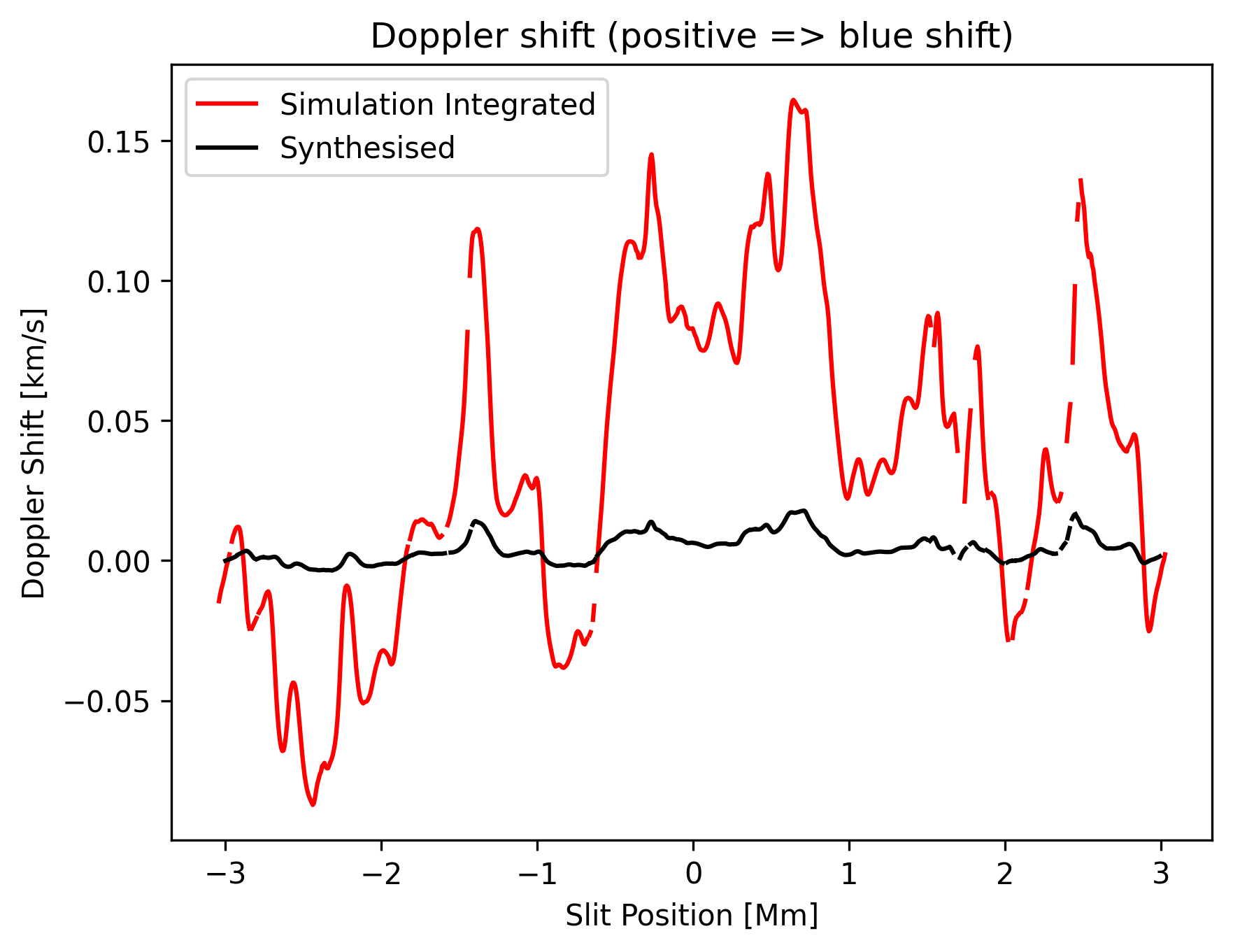}
    \includegraphics[width=0.32\linewidth]{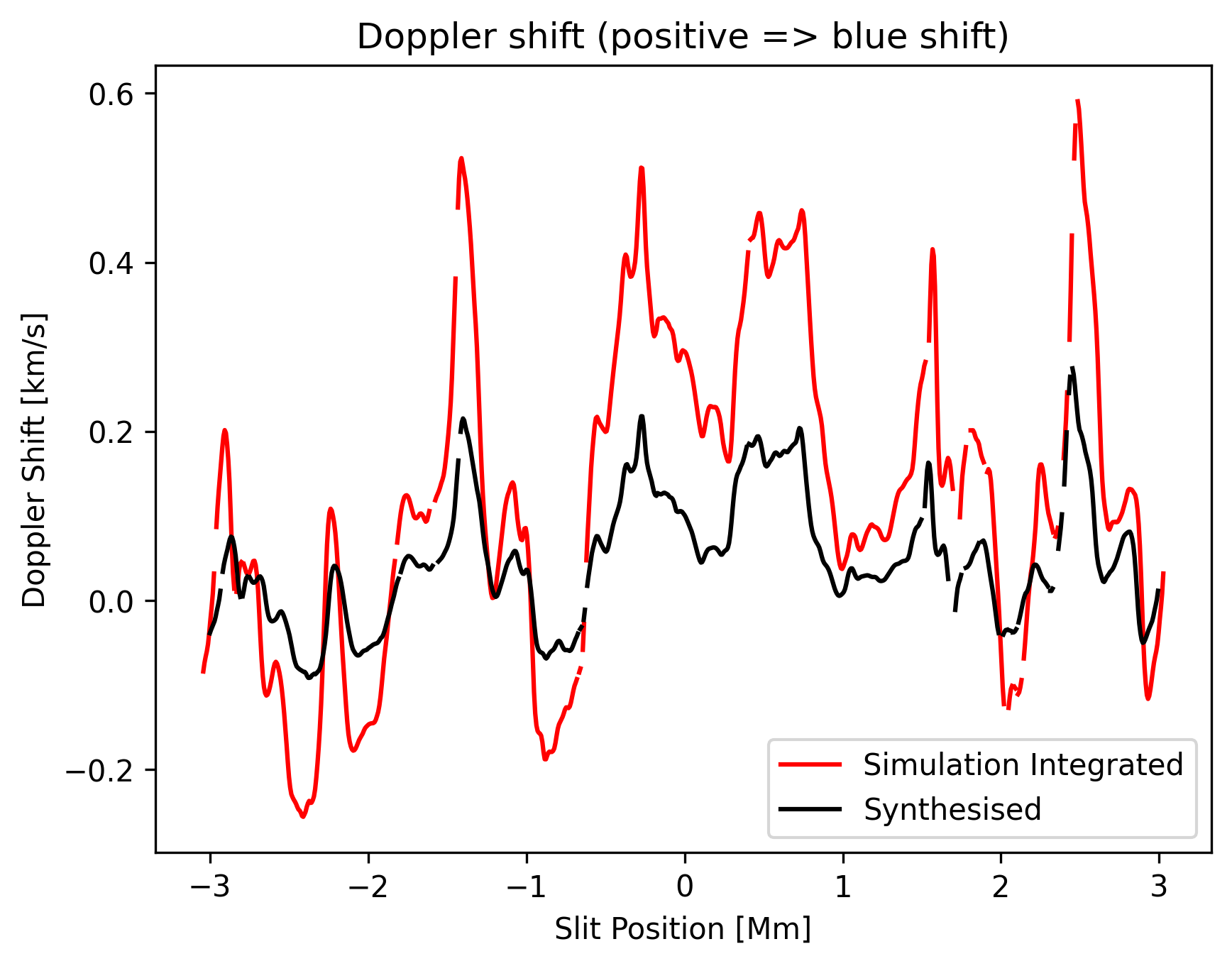}
    \includegraphics[width=0.32\linewidth]{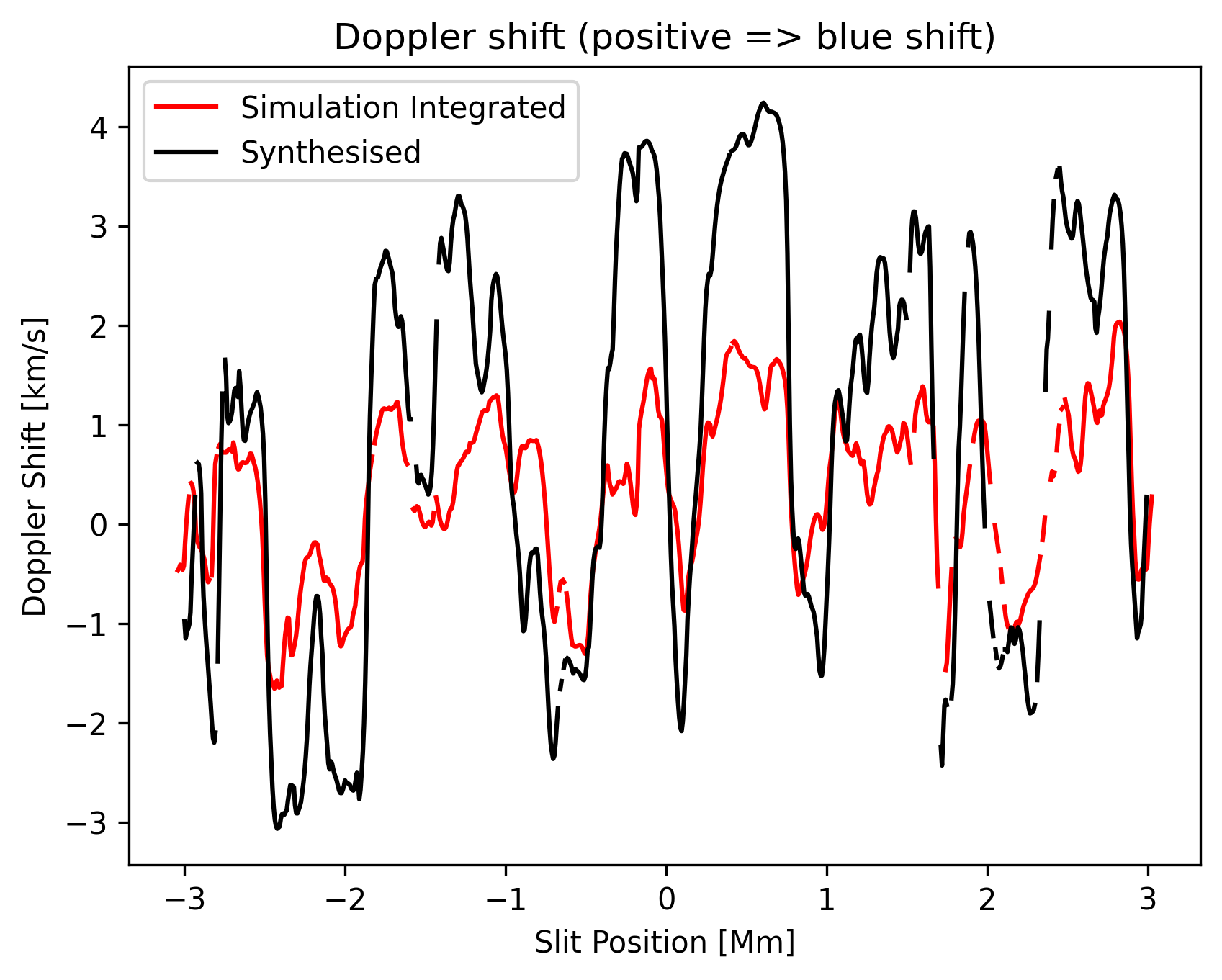}
    \caption{Doppler velocity of the synthesised spectral emission (black) from H$\alpha$ (left), Ca\,\textsc{ii}\,K (centre), Mg\,\textsc{ii}\,h (right), compared to the emission-weighted Doppler velocity from the simulation data (red) at time $\hat{t}=1200$s (corresponding to simulation time $t=200$). Unconverged data is treated as missing data.}
    \label{fig:spectraDoppler}
\end{figure*}

\section{Spectra}


The synthetic slit spectrograms from H$\alpha$, Mg\,\textsc{ii}\,h and Ca\,\textsc{ii}\,K are shown in Figure \ref{fig:slitspectra} at time $\hat{t}=1200$s (corresponding to simulation time $t=200$ in Figure \ref{fig:simevo}). The line profile is very close to Gaussian for Ca\,\textsc{ii}\,K and H$\alpha$ across the length of the simulation, with peak emission occurring at the line core and very little asymmetry in either wing, as shown in Figure \ref{fig:spectraAsym}. Here, the asymmetries are calculated following \citet{Kerr2015} and \citet{Peat2021} as:
\begin{gather}
    \mbox{Asymmetry} = \frac{(\lambda_{88}-\lambda_{50})-(\lambda_{50}-\lambda_{12})}{\lambda_{88}-\lambda_{12}}
\end{gather}
where $\lambda_{50}$ is the centre-of-mass of the line, and $\lambda_{12},\lambda_{88}$ are the 12\% and 88\% of the cumulative distribution function respectively. 
The Mg\,\textsc{ii}\,h spectra behaves differently, with a central reversal and secondary emission peaks in the blue and red wings rather than the line core.
There is also an asymmetry present, as highlighted in Figure \ref{fig:spectraAsym}. 

The Doppler velocities in each line can be calculated using the line shift as:
\begin{gather}
    v_{\rm D,synth}=c_l \frac{\lambda_{50}-\lambda}{\lambda}
\end{gather}
where $\lambda$ is the wavelength of the line core. The synthesised Doppler velocities are shown in Figure \ref{fig:spectraDoppler}. As expected from the formation locations of the spectral lines (Figure \ref{fig:combined_formation}), the Doppler velocity in the H$\alpha$ and Ca\,\textsc{ii}\,K lines is very small as these lines are not sampling a region that is undergoing substantial mixing. The Mg\,\textsc{ii}\,h line forms within the mixing layer itself and thus the Doppler velocity calculated from the line shift is on the order of a few thousand m/s and is generally blue-shifted. 

The velocity from the centroid method can be compared to the velocity in the simulation data. To obtain an estimate of the velocity that contributes towards the Doppler velocity in each line, the simulation velocity is calculated using:
\begin{gather}
    v_D=\frac{\int C v_y dl}{\int C dl}
\end{gather}
where $C$ is the intensity contribution function for a given line, as shown in Figure \ref{fig:combined_formation}, and $v_y$ is the component of velocity aligned with the line-of-sight. The weighted velocity $v_D$ therefore comes only from the velocity that is in the regions where the line forms. For H$\alpha$, the Doppler velocity estimated from the centroid method is much smaller than the value directly from the simulation data, however, the magnitudes are small. In Ca\,\textsc{ii}\,K, the shape of the synthesised and simulation velocities pair up reasonably well, however the magnitude of the synthesised value is roughly half that of the simulation value. For Mg\,\textsc{ii}\,h, the Doppler velocity from the line shift is a factor of roughly 2 lower than the simulation velocity. The velocity magnitude for the Ca\,\textsc{ii}\,K and Mg\,\textsc{ii}\,h lines are reasonably close to the simulation value.  

The RMS speeds from the Doppler velocity from the line shift are {0.006.5 km/s (H$\alpha$), 0.091 km/s (Ca\,\textsc{ii}\,K) and 2.243 km/s (Mg\,\textsc{ii}\,h)}. The Mg\,\textsc{ii}\,h line forms primarily within the mixing layer between the thread and corona, and hence the RMS velocity can be compared to the expected mean velocity in the mixing layer. Based on the theory of \citet{Hillier2019} one would expect that the RMS velocity in the mixing layer is 
\begin{equation}
    V_{\rm RMS}\approx\sqrt{\frac{1}{2}}\frac{(\rho_{\rm thread}\rho_{\rm corona})^{1/4}}{\sqrt{\rho_{\rm thread}}+\sqrt{\rho_{\rm corona}}}\Delta V,
\end{equation}
which for the simulation setup presented in this paper is {$\approx 3.4$\,km/s}. This value is in line with the fluctuations in Doppler velocity that are inferred using the synthesised Mg\,\textsc{ii}\,h line highlighting its potential strength for probing turbulent mixing in prominences.


\section{Time series}

\begin{figure}
    \includegraphics[width=\columnwidth]{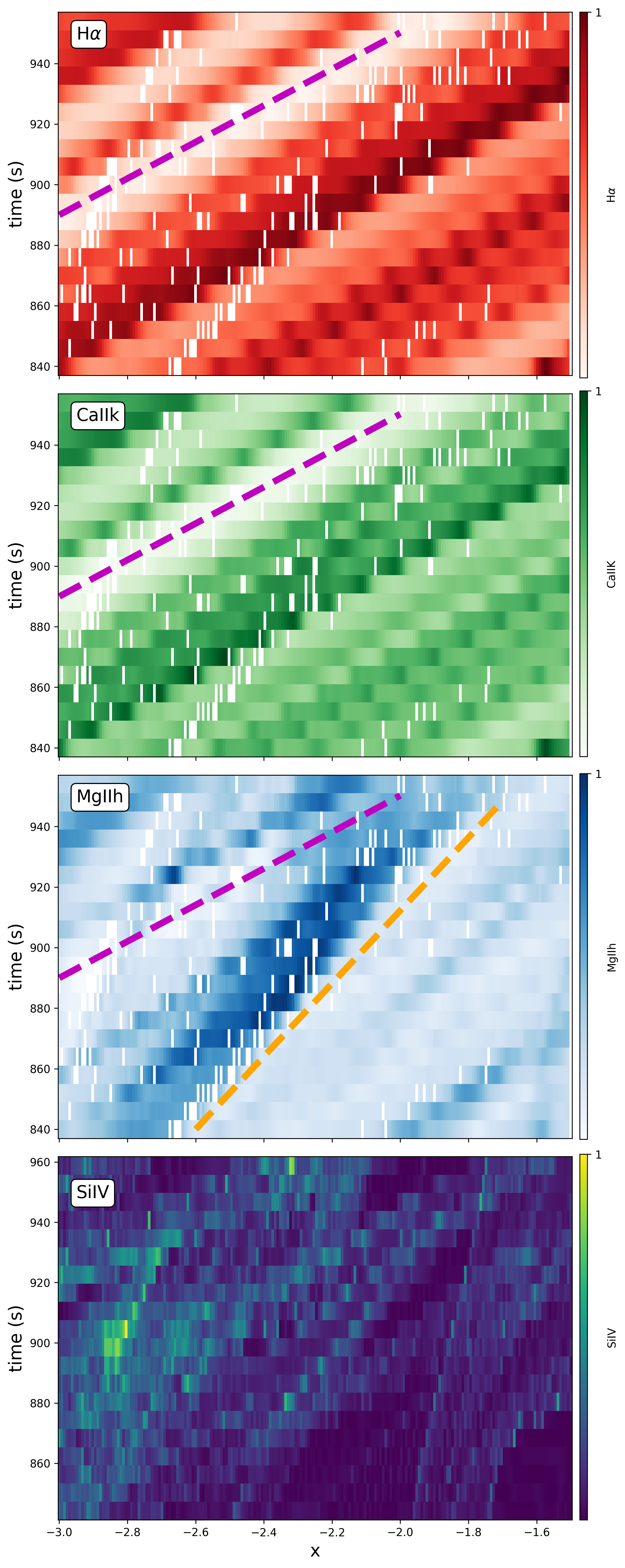}
    \caption{Normalised intensity timeseries for H$\alpha$, Ca\,\textsc{ii}\,K, Mg\,\textsc{ii}\,h and Si\,\textsc{iv} (listed from top to bottom). The white locations are where the line synthesis did not converge. Magenta dashed line indicates a speed of {16.6 km/s}. Orange dashed line is {8.3 km/s}.}
    \label{fig:intens}
\end{figure}

The synthesised emission is shown in Figure \ref{fig:intens} through time for all $x-$locations in a 1.5 Mm slit with a temporal resolution of 6 seconds (corresponding to each simulation output). A small time window of $\approx 120$ seconds is shown in Figure \ref{fig:intens}. 

From Figure \ref{fig:intens}, it is clear that different features are being sampled by the different spectral lines. H$\alpha$ and Ca\,\textsc{ii}\,K sample the thread core and edge respectively and thus similar structures are present. There is very little mixing at these locations and the features are moving at an approximately constant rate of $\approx 16.6$ km/s, as indicated by the magenta line in Figure \ref{fig:intens}, corresponding to the speed at which the thread is falling, see Table \ref{tab:constants}. This is expected given the locations sampled by these lines, shown in Figure \ref{fig:combined_formation}.

The Mg\,\textsc{ii}\,h line samples the mixing layer itself, and structures can be seen falling at different rates. Some emission appears to follow the thread (16.6 km/s), while other brighter structures fall at a rate approximately half this (8.3 km/s), which is the mean flow between the falling thread and stationary corona (indicated by the orange line in Figure \ref{fig:intens}). 

The Si\,\textsc{iv} emission shows no dominant structures. There is a range of structures that overlap and intersect. This may be due to the turbulent material sampled by Si\,\textsc{iv} being less-influenced by the bulk motion of the thread than the cooler lines. 

\begin{figure}
     \includegraphics[width=\columnwidth]{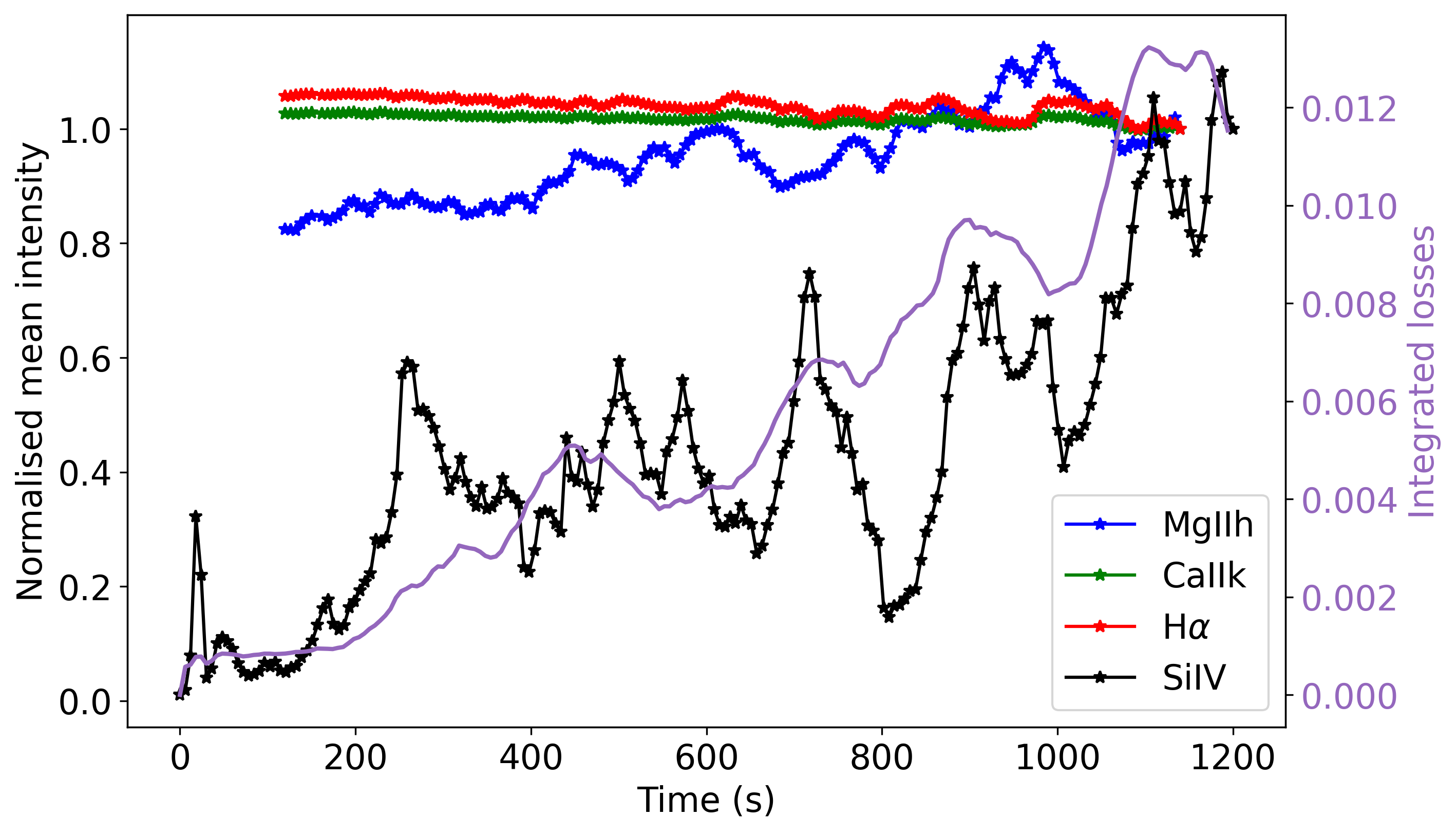}
    \caption{Mean emission along the $x-$direction though time for the Si\,\textsc{iv} (black), H$\alpha$ (red), Ca\,\textsc{ii}\,K (green) and Mg\,\textsc{ii}\,h (blue) spectral lines.}
    \label{fig:meanintens}
\end{figure}

The mean emission through time for the spectral lines is shown in Figure \ref{fig:meanintens}, with the intensity for each line normalised to unity at time 1200 s. The mean emission is calculated at each time based on the locations that converged, with convergence obtained in $\approx90\%$ of the domain on average. 
There is a small decrease in the intensity of the Ca\,\textsc{ii}\,K and H$\alpha$ lines with time. H$\alpha$ and Ca\,\textsc{ii}\,K primarily sample the thread core and a thin layer at the mixing edge respectively, as shown in Figure \ref{fig:combined_formation}a. As the thread falls, the dense core of the thread is entrained into the mixing layer and corona, thus the total amount of material decreases, hence the total emission in these lines decreases.

The mean intensity emission of Mg\,\textsc{ii}\,h varies over the time window. As material mixes, it forms intermediate temperatures that are sampled by the Mg\,\textsc{ii}\,h line, as shown in Figure \ref{fig:combined_formation}. The material in this range is rapidly cooled and thus the net emission is very sensitive to the mixing phenomena.

For the Si\,\textsc{iv} line, a strong increase in the mean emission is observed through time, that correlates with the integrated radiative losses in the domain. Observationally, an increased emission in a hotter line could be interpreted as heating. However, this explanation is inconsistent with the simulation data, where the heating is negligible compared to the cooling and there is a net {loss} of thermal energy through the simulation.


\section{Discussion}



\subsection{Doppler velocity from spectra vs simulation}

In the Ca\,\textsc{ii}\,K and MgII lines, the Doppler velocity derived from the spectra agrees well with the emission-weighted velocity from the simulation, with a difference of a factor of roughly 2. The Mg\,\textsc{ii}\,h is an over-estimate and Ca\,\textsc{ii}\,K is an under-estimate, giving some indication of the error bars that may be present in the velocity estimated from the centroid method. The velocity estimated from the centroid method for the H$\alpha$ line is significantly smaller than the simulation, however the magnitude of the velocity in this region is very small. It may be that the inherent symmetry of the H$\alpha$ line profile (shown in Figure \ref{fig:slitspectra}) prohibits accurate measurement of such small velocities with the centroid method.

The RMS velocity from the Mg\,\textsc{ii}\,h line shifts is approximately 2.2\,km/s, which can be compared to the theoretical value proposed by \cite{Hillier2019} which, for the setup presented here, is 3.4\,km/s. These values are reasonably close and thus the RMS velocity estimation of \cite{Hillier2019} has potential strengths in probing turbulent mixing layers in prominences.


\subsection{Mixing phenomena may drive {increased} emission in the absence of heating}

The observed signature of a decrease in cool emission (e.g., Ca\,\textsc{ii}\,K) and an increase in hotter emission (e.g., Si\,\textsc{iv}) is often interpreted as a signature of heating that is driven by turbulent heating. Here, the underlying simulation used in this study has strong cooling that far exceeds the turbulent heating in the simulation and leads to a net loss of thermal energy through time, and yet there is a clear increase in emission in Si\,\textsc{iv} throughout the mixing process. As such, the fundamental mixing process may be responsible for the observed increase in hotter emission, without the requirement for heating.

Only a small proportion of the thread material is lost through the simulation, $\approx 10\%$. However, even this is enough to reduce the emission in the Ca\,\textsc{ii}\,K and H$\alpha$ lines throughout the mixing process. It is likely that a longer simulation would feature more substantial mass loss from the thread, further decreasing the mean intensity in cool spectral lines.

\begin{figure}
    \includegraphics[width=\columnwidth]{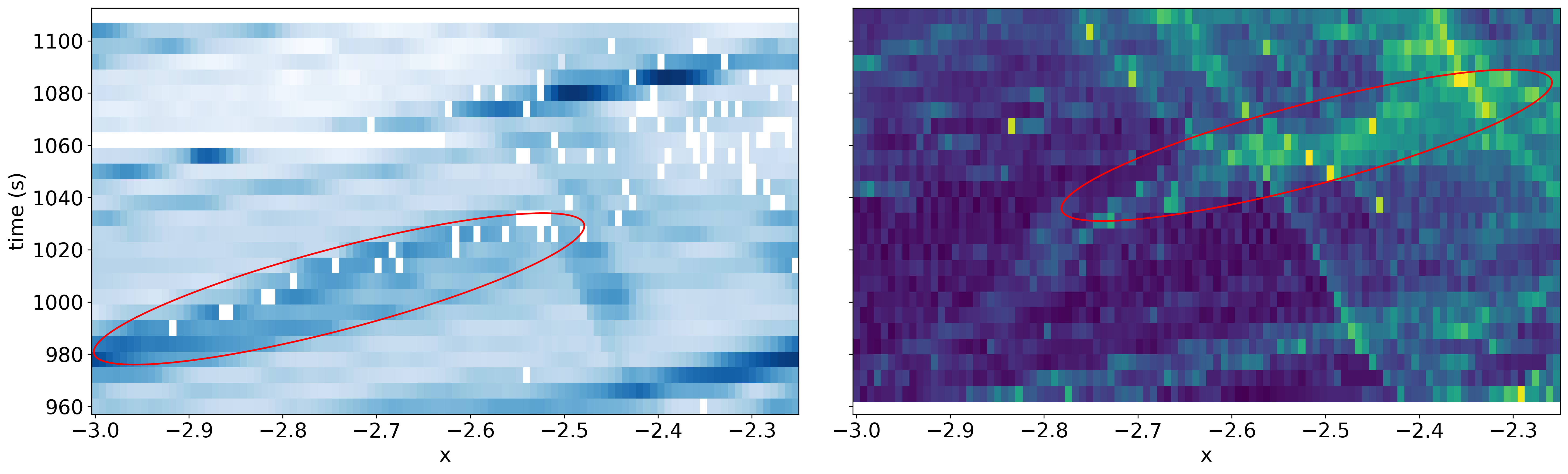}
    \caption{Optically thick intensity though time for Mg\,\textsc{ii}\,h (left) and Si\,\textsc{iv} (right) for a selected time window. The white locations are where the line synthesis did not converge.}
    \label{fig:subset_test}
\end{figure}

A selected time-window for the Mg\,\textsc{ii}\,h and Si\,\textsc{iv} emission is shown in Figure \ref{fig:subset_test}. This time window shows a feature that appears to fade from the cooler Mg\,\textsc{ii}\,h line, and reappear in the hotter Si\,\textsc{iv}. Qualitatively, this is similar to the observed result of co-temporal fading in cool lines and enhanced emission in hotter lines \citep[e.g.,][]{Okamoto2015}. However, the simulation in this paper is dominated by cooling and the feature forms at a time and location where the is entrainment of mixed material, which may be responsible for this signature.

There are many caveats to this result, most importantly, this is a very cherry-picked time window and not a consistent feature of the time series. However, it does highlight that the observed signature of fading in cool lines and enhanced emission in warm lines can arise from a simulation that is cooling and is not necessarily a signature of heating. 


\subsection{Thermal conduction}

The 2D simulation presented in this paper features an out-of-plane magnetic field and thus thermal conduction does not affect the evolution of the model. However, the solar atmosphere is 3D and thermal conduction is essential for the thermal evolution of the solar corona. Thermal conduction would act to to prevent steep temperature gradients forming in the mixing layer and thus may diffuse the temperature in the mixing layer along the direction of the magnetic field. This may result in a broader emission region from the Mg\,\textsc{ii}\,h line.

\subsection{Choice of normalisation length}

The temperature normalisation is inherently built into the simulation through the radiative losses, however the reference length and density can be chosen a-posteriori. A change of the length scale changes the time normalisation since $\hat{t}=v_{0}/L_0$.

A smaller lengthscale would result in a thinner thread which changes the opacity for the optically thick emission. As such, at small enough lengthscales, the emission may feature contributions from both sides of the thread, as opposed to the results of this paper where the bulk emission comes from the observer-side, as shown in Figure \ref{fig:combined_formation}. Synthesising the emission using $L_0=5\times 10^5$m results in the H$\alpha$ emission spanning the whole thread. However, emission is still only on the observer-side for the Mg\,\textsc{ii}\,h and Ca\,\textsc{ii}\,K spectral lines. As such, for these lines, the opposite side of the thread may only have a non-negligible contribution to emission for narrow ($<5\times 10^5$m) threads. Note that this is only true at line core and the wings of Mg\,\textsc{ii}\,h can feature contribution from both sides of the thread even in the $L_0=10^6$m case presented in this paper. 
This is in spite of the high optical depth of Mg\,\textsc{ii}\,h in the line core, and likely due to the differing Doppler shifts moving the narrow emission window.

\section{Conclusions}

The mixing between cool ($10^4$K) and hot ($10^6$K) material in the solar atmosphere creates a complex mixing region which is subject to efficient cooling due to radiative losses. In this paper, a falling prominence thread is modelled and the observational signatures of the resultant mixing is studied. The optically-thick synthesis reveals that H$\alpha$ and Ca\,\textsc{ii}\,K sample the thread and its boundary respectively, whereas the Mg\,\textsc{ii}\,h line forms during the mixing layer. 

The fundamental mixing process drives intermediate temperatures ($10^5$K) to form {within} the mixing layer, leading to decreased emission in cool lines (e.g., Ca\,\textsc{ii}\,K) an increased emission in warm lines (e.g., Si\,\textsc{iv}). However, the increased warm emission is coincided by an increased energy loss rate due to the efficient radiative losses that act in the mixing layer. The radiative losses far exceed the turbulent heating and thus an increased emission in warm lines at the prominence-corona interface may be a signature of mixing, rather than heating.



\section*{Acknowledgements}
BS and AH are supported by STFC research grant ST/V000659/1. 
CMJO is grateful for the support of a Royal Astronomical Society fellowship. This research was supported by the International Space Science Institute (ISSI) in Bern, through ISSI International Team project \#545. For the purpose of open access, the author has applied a Creative Commons Attribution (CC BY) licence to any Author Accepted Manuscript version arising from this submission’

\section*{Data Availability}

The (P\underline{I}P) numerical solver used in this study is freely available \href{https://github.com/AstroSnow/PIP}{https://github.com/AstroSnow/PIP}

The optically-thick synthesis is performed using the open-source \Lw{} framework \href{https://goobley.github.io/Lightweaver/}{https://goobley.github.io/Lightweaver/}

Data is available from BS upon reasonable request.



\bibliographystyle{mnras}
\bibliography{CoolingSignatures} 

\begin{thebibliography}{}
\makeatletter
\relax
\def\mn@urlcharsother{\let\do\@makeother \do\$\do\&\do\#\do\^\do\_\do\%\do\~}
\def\mn@doi{\begingroup\mn@urlcharsother \@ifnextchar [ {\mn@doi@}
  {\mn@doi@[]}}
\def\mn@doi@[#1]#2{\def\@tempa{#1}\ifx\@tempa\@empty \href
  {http://dx.doi.org/#2} {doi:#2}\else \href {http://dx.doi.org/#2} {#1}\fi
  \endgroup}
\def\mn@eprint#1#2{\mn@eprint@#1:#2::\@nil}
\def\mn@eprint@arXiv#1{\href {http://arxiv.org/abs/#1} {{\tt arXiv:#1}}}
\def\mn@eprint@dblp#1{\href {http://dblp.uni-trier.de/rec/bibtex/#1.xml}
  {dblp:#1}}
\def\mn@eprint@#1:#2:#3:#4\@nil{\def\@tempa {#1}\def\@tempb {#2}\def\@tempc
  {#3}\ifx \@tempc \@empty \let \@tempc \@tempb \let \@tempb \@tempa \fi \ifx
  \@tempb \@empty \def\@tempb {arXiv}\fi \@ifundefined
  {mn@eprint@\@tempb}{\@tempb:\@tempc}{\expandafter \expandafter \csname
  mn@eprint@\@tempb\endcsname \expandafter{\@tempc}}}

\bibitem[\protect\citeauthoryear{{Antolin}}{{Antolin}}{2020}]{Antolin2020}
{Antolin} P.,  2020, \mn@doi [Plasma Physics and Controlled Fusion]
  {10.1088/1361-6587/ab5406}, \href
  {https://ui.adsabs.harvard.edu/abs/2020PPCF...62a4016A} {62, 014016}

\bibitem[\protect\citeauthoryear{{Antolin}, {Okamoto}, {De Pontieu},
  {Uitenbroek}, {Van Doorsselaere}  \& {Yokoyama}}{{Antolin}
  et~al.}{2015}]{Antolin2015}
{Antolin} P.,  {Okamoto} T.~J.,  {De Pontieu} B.,  {Uitenbroek} H.,  {Van
  Doorsselaere} T.,   {Yokoyama} T.,  2015, \mn@doi [\apj]
  {10.1088/0004-637X/809/1/72}, \href
  {https://ui.adsabs.harvard.edu/abs/2015ApJ...809...72A} {809, 72}

\bibitem[\protect\citeauthoryear{{Beckers}}{{Beckers}}{1968}]{Beckers1968}
{Beckers} J.~M.,  1968, \mn@doi [\solphys] {10.1007/BF00171614}, \href
  {https://ui.adsabs.harvard.edu/abs/1968SoPh....3..367B} {3, 367}

\bibitem[\protect\citeauthoryear{{Beckers}}{{Beckers}}{1972}]{Beckers1972}
{Beckers} J.~M.,  1972, \mn@doi [\araa] {10.1146/annurev.aa.10.090172.000445},
  \href {https://ui.adsabs.harvard.edu/abs/1972ARA&A..10...73B} {10, 73}

\bibitem[\protect\citeauthoryear{{Dere}, {Landi}, {Mason}, {Monsignori Fossi}
  \& {Young}}{{Dere} et~al.}{1997}]{Dere1997}
{Dere} K.~P.,  {Landi} E.,  {Mason} H.~E.,  {Monsignori Fossi} B.~C.,   {Young}
  P.~R.,  1997, \mn@doi [\aaps] {10.1051/aas:1997368}, \href
  {https://ui.adsabs.harvard.edu/abs/1997A&AS..125..149D} {125, 149}

\bibitem[\protect\citeauthoryear{{Dere}, {Del Zanna}, {Young}, {Landi}  \&
  {Sutherland}}{{Dere} et~al.}{2019}]{Dere2019}
{Dere} K.~P.,  {Del Zanna} G.,  {Young} P.~R.,  {Landi} E.,   {Sutherland}
  R.~S.,  2019, \mn@doi [\apjs] {10.3847/1538-4365/ab05cf}, \href
  {https://ui.adsabs.harvard.edu/abs/2019ApJS..241...22D} {241, 22}

\bibitem[\protect\citeauthoryear{{Fontenla}, {Avrett}  \& {Loeser}}{{Fontenla}
  et~al.}{1993}]{Fontenla1993}
{Fontenla} J.~M.,  {Avrett} E.~H.,   {Loeser} R.,  1993, \mn@doi [\apj]
  {10.1086/172443}, \href
  {https://ui.adsabs.harvard.edu/abs/1993ApJ...406..319F} {406, 319}

\bibitem[\protect\citeauthoryear{{Heinzel}, {Schmieder}, {Mein}  \&
  {Gun{\'a}r}}{{Heinzel} et~al.}{2015}]{Heinzel2015}
{Heinzel} P.,  {Schmieder} B.,  {Mein} N.,   {Gun{\'a}r} S.,  2015, \mn@doi
  [\apjl] {10.1088/2041-8205/800/1/L13}, \href
  {https://ui.adsabs.harvard.edu/abs/2015ApJ...800L..13H} {800, L13}

\bibitem[\protect\citeauthoryear{{Hillier} \& {Arregui}}{{Hillier} \&
  {Arregui}}{2019}]{Hillier2019}
{Hillier} A.,  {Arregui} I.,  2019, \mn@doi [\apj] {10.3847/1538-4357/ab4795},
  \href {https://ui.adsabs.harvard.edu/abs/2019ApJ...885..101H} {885, 101}

\bibitem[\protect\citeauthoryear{{Hillier}, {Takasao}  \& {Nakamura}}{{Hillier}
  et~al.}{2016}]{Hillier2016}
{Hillier} A.,  {Takasao} S.,   {Nakamura} N.,  2016, \mn@doi [\aap]
  {10.1051/0004-6361/201628215}, \href
  {https://ui.adsabs.harvard.edu/abs/2016A&A...591A.112H} {591, A112}

\bibitem[\protect\citeauthoryear{{Hillier}, {Snow}  \& {Arregui}}{{Hillier}
  et~al.}{2023}]{Hillier2023}
{Hillier} A.,  {Snow} B.,   {Arregui} I.,  2023, \mn@doi [\mnras]
  {10.1093/mnras/stad234}, \href
  {https://ui.adsabs.harvard.edu/abs/2023MNRAS.520.1738H} {520, 1738}

\bibitem[\protect\citeauthoryear{Hinode Review~Team et~al.,}{Hinode Review~Team
  et~al.}{2019}]{Hinode2019}
Hinode Review~Team a.,  et~al., 2019, \mn@doi [Publications of the Astronomical
  Society of Japan] {10.1093/pasj/psz084}, 71, R1

\bibitem[\protect\citeauthoryear{{Jenkins}, {Osborne}  \& {Keppens}}{{Jenkins}
  et~al.}{2023}]{Jenkins2023}
{Jenkins} J.~M.,  {Osborne} C.~M.~J.,   {Keppens} R.,  2023, \mn@doi [\aap]
  {10.1051/0004-6361/202244868}, \href
  {https://ui.adsabs.harvard.edu/abs/2023A&A...670A.179J} {670, A179}

\bibitem[\protect\citeauthoryear{{Kerr}, {Sim{\~o}es}, {Qiu}  \&
  {Fletcher}}{{Kerr} et~al.}{2015}]{Kerr2015}
{Kerr} G.~S.,  {Sim{\~o}es} P.~J.~A.,  {Qiu} J.,   {Fletcher} L.,  2015,
  \mn@doi [\aap] {10.1051/0004-6361/201526128}, \href
  {https://ui.adsabs.harvard.edu/abs/2015A&A...582A..50K} {582, A50}

\bibitem[\protect\citeauthoryear{{Labrosse} \& {Gouttebroze}}{{Labrosse} \&
  {Gouttebroze}}{2004}]{Labrosse2004}
{Labrosse} N.,  {Gouttebroze} P.,  2004, \mn@doi [\apj] {10.1086/425168}, \href
  {https://ui.adsabs.harvard.edu/abs/2004ApJ...617..614L} {617, 614}

\bibitem[\protect\citeauthoryear{{Leenaarts}, {Pereira}  \&
  {Uitenbroek}}{{Leenaarts} et~al.}{2012}]{Leenaarts2012}
{Leenaarts} J.,  {Pereira} T.,   {Uitenbroek} H.,  2012, \mn@doi [\aap]
  {10.1051/0004-6361/201219394}, \href
  {https://ui.adsabs.harvard.edu/abs/2012A&A...543A.109L} {543, A109}

\bibitem[\protect\citeauthoryear{{Okamoto}, {Antolin}, {De Pontieu},
  {Uitenbroek}, {Van Doorsselaere}  \& {Yokoyama}}{{Okamoto}
  et~al.}{2015}]{Okamoto2015}
{Okamoto} T.~J.,  {Antolin} P.,  {De Pontieu} B.,  {Uitenbroek} H.,  {Van
  Doorsselaere} T.,   {Yokoyama} T.,  2015, \mn@doi [\apj]
  {10.1088/0004-637X/809/1/71}, \href
  {https://ui.adsabs.harvard.edu/abs/2015ApJ...809...71O} {809, 71}

\bibitem[\protect\citeauthoryear{{Osborne} \& {Mili{\'c}}}{{Osborne} \&
  {Mili{\'c}}}{2021}]{Osborne2021}
{Osborne} C. M.~J.,  {Mili{\'c}} I.,  2021, \mn@doi [\apj]
  {10.3847/1538-4357/ac02be}, \href
  {https://ui.adsabs.harvard.edu/abs/2021ApJ...917...14O} {917, 14}

\bibitem[\protect\citeauthoryear{{Parenti}}{{Parenti}}{2014}]{Parenti2014}
{Parenti} S.,  2014, \mn@doi [Living Reviews in Solar Physics]
  {10.12942/lrsp-2014-1}, \href
  {https://ui.adsabs.harvard.edu/abs/2014LRSP...11....1P} {11, 1}

\bibitem[\protect\citeauthoryear{{Peat}, {Labrosse}, {Schmieder}  \&
  {Barczynski}}{{Peat} et~al.}{2021}]{Peat2021}
{Peat} A.~W.,  {Labrosse} N.,  {Schmieder} B.,   {Barczynski} K.,  2021,
  \mn@doi [\aap] {10.1051/0004-6361/202140907}, \href
  {https://ui.adsabs.harvard.edu/abs/2021A&A...653A...5P} {653, A5}

\bibitem[\protect\citeauthoryear{{Pereira}}{{Pereira}}{2019}]{Pereira2019}
{Pereira} T. M.~D.,  2019, \mn@doi [Advances in Space Research]
  {10.1016/j.asr.2018.09.030}, \href
  {https://ui.adsabs.harvard.edu/abs/2019AdSpR..63.1434P} {63, 1434}

\bibitem[\protect\citeauthoryear{{Pereira} et~al.,}{{Pereira}
  et~al.}{2014}]{Pereira2014}
{Pereira} T.~M.~D.,  et~al., 2014, \mn@doi [\apjl]
  {10.1088/2041-8205/792/1/L15}, \href
  {https://ui.adsabs.harvard.edu/abs/2014ApJ...792L..15P} {792, L15}

\bibitem[\protect\citeauthoryear{{Priest}}{{Priest}}{1982}]{Priest1982}
{Priest} E.~R.,  1982, {Solar magneto-hydrodynamics.}.
 Vol. 21, Springer Dordrecht

\bibitem[\protect\citeauthoryear{{Rybicki} \& {Hummer}}{{Rybicki} \&
  {Hummer}}{1992}]{Rybicki1992}
{Rybicki} G.~B.,  {Hummer} D.~G.,  1992, \aap, \href
  {https://ui.adsabs.harvard.edu/abs/1992A&A...262..209R} {262, 209}

\bibitem[\protect\citeauthoryear{{Scott} et~al.,}{{Scott}
  et~al.}{2015}]{Scott2015}
{Scott} P.,  et~al., 2015, \mn@doi [\aap] {10.1051/0004-6361/201424109}, \href
  {https://ui.adsabs.harvard.edu/abs/2015A&A...573A..25S} {573, A25}

\bibitem[\protect\citeauthoryear{{Uitenbroek}}{{Uitenbroek}}{2001}]{Uitenbroek2001}
{Uitenbroek} H.,  2001, \mn@doi [\apj] {10.1086/321659}, \href
  {https://ui.adsabs.harvard.edu/abs/2001ApJ...557..389U} {557, 389}

\bibitem[\protect\citeauthoryear{{de Pontieu} et~al.,}{{de Pontieu}
  et~al.}{2007}]{dePontieu2007}
{de Pontieu} B.,  et~al., 2007, \mn@doi [\pasj] {10.1093/pasj/59.sp3.S655},
  \href {https://ui.adsabs.harvard.edu/abs/2007PASJ...59S.655D} {59, S655}

\makeatother
\end{thebibliography}




\appendix

\bsp	
\label{lastpage}
\end{document}